\documentclass[aps,
superscriptaddress]{revtex4}
\usepackage[intlimits]{amsmath}
\usepackage{amsfonts}
\usepackage{graphics}

\usepackage{amssymb}
\usepackage[usenames]{color}
\usepackage{epstopdf,epsfig}
\usepackage{subfigure}
\usepackage{indentfirst}
\usepackage{graphicx}
\usepackage{amsthm}

\theoremstyle{plain}
\newtheorem{thm}{Theorem}
\newtheorem{lem}{Lemma}
\newtheorem{Def}{Definition}
\newtheorem{cor}{Corollary}
\newtheorem{pro}{Proposition}

\newtheorem*{rem*}{Remark}
\newtheorem*{def*}{Definition}
\newtheorem*{prob*}{Direct Monodromy Problem}
\newtheorem*{lem*}{\textsc{Lemma}}
\newtheorem*{cor*}{\textsc{Corollary}}
\newtheorem*{con*}{\textsc{Conjecture}}
\newcommand{\e}{\varepsilon}
\newcommand{\bb}[1]{\mathbb{ #1 }}
\newcommand{\ma}[2]{\lbrace #1 , #2 \rbrace}

\newcommand{\beq}{\begin{equation}}
\newcommand{\eeq}{\end{equation}}

\begin{document}
\title{A Laplace's method for series \\ and the semiclassical analysis of epidemiological models}
\author{Davide Masoero}\email{dmasoero@gmail.com}\affiliation{Grupo de F\'isica Matem\'atica da Universidade de Lisboa}

\begin{abstract}
 We develop a Laplace's method to compute the asymptotic expansions of sums of sharply peaked sequences.
These series arise as discretizations (Riemann sums) of sharply-peaked  integrals,
whose asymptotic behavior can be computed by the standard Laplace's method.
We apply the Laplace's method for series to the WKB (i.e. semiclassical) analysis of stochastic models of population biology,
with special focus on the SIS model.
In particular we show that two different and widely-used approaches to the semiclassical limit, i.e. either considering a semiclassical
probability distribution or a semiclassical generating function, are equivalent. 

\end{abstract}

\maketitle

\section*{Introduction}

In this paper we develop a Laplace's method to deal with series whose summand is sharply peaked about its maximum.
More precisely, the main object of the paper is the asymptotic evaluation of the series
\begin{equation}\label{eq:INalpha}
 I(n,\alpha)= n^{1-\alpha} \sum_{k=0}^{\infty} e^{-n f(\frac{k}{n^{\alpha}})}g(\frac{k}{n^{\alpha}}) \, , 
 \alpha>0  \, , n \to + \infty \;,
\end{equation}
with the particular aim of providing a mathematical foundation to the WKB (semiclassical) analysis of epidemiological models.

The series $I(n,\alpha)$ is actually a Riemann sum, with respect to a homogeneous
partition of the positive semi-axis into subintervals of length $1/n^{\alpha}$, of the integral
\begin{equation}\label{eq:intN}
 {\cal{I}}(n)=n \int_0^{\infty}e^{-n f(x)}g(x) dx \; ,
\end{equation}
and the asymptotic behavior of the latter integral can be computed by the (standard) Laplace's method provided it applies, see e.g.
\cite{deburijn81} \cite{bender99} \cite{peter06}.

One can expect the series to have the same asymptotic behavior as that of the integral if the partition is fine enough, that is for $\alpha$ big,
while under a certain threshold the effects of the discretization cannot be neglected. This is indeed the case. Depending on the
local nature of the global minimum
of $f$, there is an $\alpha^*$ below which the asymptotics of the series and of the integral do not coincide, and the series is oscillatory and even
exponentially small with respect to the integral - a pictorial representation of this phenomenon can be found in Figures \ref{fig:I25} and
\ref{fig:Iunterzo} at
the end of Section
\ref{sec:laplaces}.

For example, we will show that if $f$ has a single global minimum point
$x_0 \neq 0$ such that $f''(x_0) \neq 0$ and $g(x_0)\neq 0$ then (we refer to Theorem \ref{thm:laplacealpha} and Corollary
\ref{cor:fullasymptotic} for the precise hypotheses and statement)
\begin{itemize}
 \item if $\alpha>\frac{1}{2}$,
 $$I(n,\alpha) \sim {\cal{I}}(n) \sim e^{-nf(x_0)} g(x_0)  \sqrt{\frac{2 \pi n}{f''(x_0)} } \; .$$
 \item if $\alpha=\frac{1}{2}$,
 $$I(n,\frac{1}{2}) \sim e^{-nf(x_0)} g(x_0) \sqrt{\frac{2\pi n}{f''(x_0)}}\theta_3(-\sqrt{n} \pi  x_0,e^{-\frac{2\pi^2}{f''(x_0)}}) \; , $$ where
 $\theta_3(z,q)$ is the third Jacobi theta-function \cite{bateman2}
 \item if $\alpha<\frac{1}{2}$,
 $$I(n,\alpha)  \sim n^{1-\alpha} e^{-nf(x_0)} g(x_0) \left(
 e^{-\frac{f''(x_0)}{2} n^{1-2\alpha}t^2(n^{\alpha}x_0)}+e^{-\frac{f''(x_0)}{2} n^{1-2\alpha}\big(1-t(n^{\alpha}x_0)\big)^2} \right) \; ,$$ 
 where $t(x)=\min\lbrace x-\lfloor x \rfloor, \lceil x \rceil -x \rbrace \; $ is the positive triangular wave of height $\frac{1}{2}$
and period $1$.
\end{itemize}

The reader may wonder whether such a seemingly natural object like $I(n,\alpha)$ had not already been studied and well-understood
in the literature. Contrary to our expectations we found few mathematical works devoted to it, the most recent being \cite{wong07} \cite{paris11}.
Paper \cite{wong07} deals with the case $\alpha=\frac{1}{2}$ (in our notation) and its application to \textit{q-polynomials}
while \cite{paris11} is devoted to the study of a class of hypergeometric functions defined by series that
fall in the case $\alpha=1$. The thorough discussion of the existing literature contained in \cite{paris11} shows that
the series $I(n,\alpha)$ was never considered in the general and simple form we do here.

Contrary to the existing works on the subject, our interest does not stem from the asymptotic theory of special functions, but
from a branch of applied mathematics in rapid evolution, the semiclassical limit of continuous time Markov process,
in particular processes related to 
epidemiological models such as SIS and SIRS models \cite{nico13}. 
The semiclassical limit of continuous time Markov process originated from the paper \cite{gang87}, where it was noticed
that many models of population biology (or more in general skip-free continuous time Markov chains)
can be solved by a
method similar to the WKB method for quantum mechanics in the \textit{semiclassical} asymptotic regime, see
\cite{kamenev04}, \cite{kessler07}, \cite{kamenev08}, \cite{meerson10}, \cite{nico13} for more recent developments.

The semiclassical regime arises when we consider a population of large size $n \in \bb{N}$ and we
assume that either the probability distribution $p_k$ is semiclassical, i.e.
\begin{equation*}
 p_k \sim c_n e^{-n S(\frac{k}{n})}L(\frac{k}{n}) \, , \; k \in \lbrace 0,1,\dots, n \rbrace
\end{equation*}
for some $S,L:[0,1] \to \bb{R}$
or the generating function is semiclassical, i.e.
\begin{equation*}
\Gamma(z) \equiv \sum_{k=0}^n p(k,n)z^k \sim d_n e^{n \Sigma(z)} \Lambda(z) \, , \; z>0 \; 
\end{equation*}
for some $\Sigma,\Lambda:[0,\infty[ \to \bb{R} $.
If one of the two hypotheses holds then, accordingly, either $S,L$ or $\Sigma,\Lambda$ evolve according to a pair of
PDEs of the classical mechanics, a Hamilton-Jacobi equation
and a related transport equation. For example, in the SIS model we consider below,
these pairs of equations are either
equations (\ref{eq:HJ},\ref{eq:transport}) or equations (\ref{eq:genHJ},\ref{eq:gentransport}).

Both approaches have been used with success and it turns out that the generating function
plays the role of the momentum representation in quantum mechanics \cite{gang87},\cite{kamenev04}
\cite{nico13}.
However, it was not at all clear whether the two semiclassical asymptotic regimes coincide, whether a semiclassical probability distribution
implies or not a semiclassical generating function.

We will be able to answer this and other related questions analyzing the generating function by means of the Laplace's method we will have developed.
In fact, if $z$ is real and positive then the generating function is a series of the kind $I(n,\alpha=1)$, namely
$$
\Gamma(z,n)=\sum_{k=0}^{\infty}e^{-n f(\frac kn)}g(\frac kn) \, , \; f(x)=S(x)-x \ln z \, , \, g(x)=L(x)\chi_{[0,1]} \; ,
$$
where $\chi_{[0,1]}$ is the characteristic function of the unit interval.

The Laplace's method  will then allow us to  show that if the probability
distribution is semiclassical then the generating function is semiclassical too. It will moreover
allow us to compute explicitly $\Sigma$ and $\Lambda$. For example $\Sigma$ turns out to be the
(restricted) Legendre-Fenchel transform of $S$, namely $\Sigma(z)=\sup_{x \in [0,1]}\lbrace -S(x) + x \ln z \rbrace$.
Among the various consequences of these computations - see Section \ref{section:sis} below -
the most important is probably
the equivalence, under some reasonable assumptions, of
the two approaches to the semiclassical limit of  SIS model. This situation should be compared with the analogous situation
in Quantum Mechanics: in Quantum Mechanics the wave functions in the momentum and position representations are related by the
Fourier transform,
and in the semiclassical limit the phases of the two wave functions
are one the Legendre transform of the other \cite{maslov81}.

The paper is divided into two main Sections. In Section \ref{sec:laplaces} we develop the Laplace's method for the series $I(n,\alpha)$ while
Section \ref{section:sis} is devoted to the application of Laplace's method (in the case $\alpha=1$) to the semiclassical limit of the SIS model.
In Section \ref{section:sis} we assume that the reader has some knowledge of the classical theory of Hamilton-Jacobi equation.

For the benefit of the reader, we end this Introduction with a summary of our main results about the asymptotic
behavior of $I(n,\alpha)$.

\paragraph*{Summary of the Asymptotic Behaviors of $I(n,\alpha)$}
As it is customary in the Laplace's method for integrals, and essentially without losing in generality, we suppose
that the function $f$ has a single global minimum for  $x \in [0,\infty[$ and we identify two different cases,
when the minimum point belongs to the open interval $]0,\infty[$
and when the minimum point belongs to the boundary, i.e it is $0$.

\paragraph*{Global minimum in the open interval} Let us assume that $I(n,\alpha)$ converges absolutely for $n$ big enough and
that at the minimum point $x_0 \neq 0$ the second derivative does not vanish
$f''(x_0)>0$, then
\begin{eqnarray}\label{eq:introlaplace}
I(n,\alpha) \sim e^{-nf(x_0)} g(x_0)
\left\lbrace 
\begin{matrix}
 \sqrt{\frac{2 \pi n}{f''(x_0)} } \, , \; \mbox{ if } \alpha >\frac{1}{2} \\
 \sqrt{\frac{2\pi n}{f''(x_0)}}\theta_3(-\sqrt{n} \pi  x_0,e^{-\frac{2\pi^2}{f''(x_0)}}))\, , \; \mbox{ if } \alpha=\frac{1}{2}\\
 n^{1-\alpha}  \left(
 e^{-\frac{f''(x_0)}{2} n^{1-2\alpha}t^2(n^{\alpha}x_0)}+e^{-\frac{f''(x_0)}{2} n^{1-2\alpha}\big(1-t(n^{\alpha}x_0)\big)^2} \right) \, , \; 
 \mbox{ if } 0<\alpha<\frac{1}{2} 
\end{matrix} \right.
\end{eqnarray}
where  $\theta_3(z,q)$ is the third Jacobi theta-function \cite{bateman2} and
$t(x)=\min\lbrace x-\lfloor x \rfloor, \lceil x \rceil -x \rbrace \; $.

With our assumption on $f,g$, the asymptotic behavior of the integral (\ref{eq:intN}) can be evaluated via the
standard Laplace's  \cite{peter06}, after which
${\cal{I}}(n) \sim  e^{-nf(x_0)} g(x_0) \sqrt{\frac{2 n \pi}{f''(x_0)}}$. Therefore
after Theorem \ref{thm:laplacealpha} we conclude that $I(n,\alpha) \sim {\cal{I}}(n) $ if and only if $\alpha>\frac{1}{2}$,
while if $\alpha<\frac{1}{2}$ then
$I(n,\alpha)$ oscillates as a consequence of the coarser discretization.

Specializing to the case $\alpha=1$, in Theorem \ref{thm:laplace1} we prove that $I(n,1)$ admits for (locally) smooth $f,g$ an asymptotic expansion
in odd powers of $n^{-\frac{1}{2}}$ which
coincides term by term with the asymptotic expansion of the integral ${\cal{I}}(n) $.

\paragraph*{Global minimum attained at the boundary} Assuming the global minimum is attained at $0$ and
$f'(0)>0, g(0) \neq 0$  then
\begin{eqnarray}\label{eq:introwatson}
I(n,\alpha) \sim e^{-nf(0)} g(0)
\left\lbrace 
\begin{matrix}
\frac{1}{f'(0)}  \, , \; \mbox{ if } \alpha > 1 \\
 \frac{1}{1-e^{-f'(0) }} \, , \; \mbox{ if } \alpha=1\\
 n^{1-\alpha}  \, , \; \mbox{ if } 0<\alpha<1 
\end{matrix} \right.
\end{eqnarray}
Considering that by Watson's Lemma $ {\cal{I}}(n) \sim \frac{e^{-nf(0)} g(0)}{f'(0)} $ ,\cite{peter06}, then in this case
the series is asymptotic to the integral, i.e. $I(n,\alpha)\sim {\cal{I}}(n)$, if and only if $\alpha>1$.

Specializing to $\alpha=1$, we prove in Theorem \ref{thm:Watson1} that $I(n)$ admits for (locally) smooth $f,g$ an asymptotic expansion
in powers of $n^{-1}$.

\emph{Acknowledgments}
We thank P. Miller, A. Raimondo, M. Souza and N. Stollenwerk for fruitful discussions.
The work is supported by the FCT Post Doc Fellowship number SFRH/BPD/75908/2011.

\section{Laplace's method for series}\label{sec:laplaces}
We begin our discussion of the Laplace's method for series by selecting a classes of sufficiently tame functions $f,g$ for which
the series $I(n,\alpha)$ converge for all $\alpha$.
\begin{Def}
 The ordered pair of functions $(f,g)$
 \begin{align*}
  &f:[0,\infty[ \to \bb{R} \cup \lbrace +\infty \rbrace \\
  &g:[0,\infty[ \to \bb{R}
 \end{align*}
is called admissible if the following two conditions hold:
\begin{itemize}
 \item[(i)] $f$ is bounded from below
 \item[(ii)] for every $\alpha>0$, there exist $C,M,m_0 >0$ such that 
 such that for all $m \geq m_0$ 
 \begin{equation}\label{eq:boundadmissible}
 |\sum_{k=0}^{\infty}e^{-m F(k/n^{\alpha})}g(k/n^{\alpha})|\leq C  n^{M} e^{- m \bar{f} } \; , \quad \bar{f}= \inf_{x \in [0,\infty[}f(x) \; .
 \end{equation}
\end{itemize}

\end{Def}
From later on, we will always assume that the functions $f,g$ defining $I(n,\alpha)$ are admissible.
In Lemma \ref{lem:criterion} we give a simple criterion for a pair of functions $(f,g)$ to be admissible.
This criterion shows that the conditions
on $(f,g)$ are quite mild.

 \begin{lem}\label{lem:criterion}
  If $f:[0,\infty[ \to \bb{R} \cup +\infty$ is bounded from below and it satisfies the inequality
  $$
  f(x)\geq c \log x \mbox{ for some } c > 0 \, , \; \mbox{ for } x \mbox{ big enough} \; ,
  $$
  then for any bounded $g:[0,\infty [ \to \bb{R}$, the pair $(f,g)$ is an admissible pair.
  \begin{proof}
  First notice that since $g$ is bounded, the thesis follows if we prove
 (\ref{eq:boundadmissible}) for $g$ the constant function $1$.
 
 Now we define
  $$F(x)=\min \lbrace f(x) , c \log x \rbrace \; .$$
  By construction $f(x)\geq F(x)$ and $\inf f = \inf F$, therefore
  the Lemma is proved if we show that the bound (\ref{eq:boundadmissible}) holds for the pair $(F,1)$.
   
  Now, let $[\tilde{x},\infty[$ be an half-infinite interval where $F$ coincides with $c\log x$.  We can split the series in two
   subseries for which the bound holds
 $$
 \sum_{k=0}^{\infty}e^{-m F(k/n^{\alpha})} = \sum_{k=0}^{\lfloor n^{\alpha}\tilde{x}\rfloor} e^{-m F(k/n^{\alpha})}+
 \sum_{k=\lfloor n^{\alpha}\tilde{x}\rfloor+1}^{\infty} e^{-m F(k/n^{\alpha})}
 $$
 The first sum is finite, it contains  $\lfloor n^{\alpha}\tilde{x}\rfloor +1 $ terms; all of them are bounded 
 by $e^{-m \bar{F}}$. This subseries is thus bounded by $(n^{\alpha}\tilde{x}+2)e^{-m \bar{F}}$.
 
 Let us now analyze the second subseries. Fixing $m_0$ to be any natural number bigger than $1/c$, we get
   \begin{align*}
& \sum_{k=\lfloor n^{\alpha}\tilde{x}\rfloor+1}^{\infty} e^{-m F(k/n^{\alpha})} \leq  e^{-(m-m_0)\bar{F}}
 \sum_{k=\lfloor n^{\alpha}\tilde{x}\rfloor+1}^{\infty}e^{-m_0 F(k/n^{\alpha})} \leq 
 \kappa n^{c \alpha n_0} e^{-m\bar{F}} \;  ,
\end{align*}
where $\kappa=e^{m_0\bar{F}}\sum_{k=\lfloor \tilde{x}\rfloor+1}^{\infty}k^{-m_0 c}$.
The thesis is proven.
  \end{proof}

 \end{lem}

The peculiar property of the series $I(n,\alpha)$ is that the sequence $e^{-nf(k/n^{\alpha)}}g(k/n^{\alpha})$ is sharply peaked
around $k \sim n^{\alpha x_0}$, where $x_0$ is the global minimum point of the function $f$. As a consequence,
even though the series depends on every term of the sequence,
its asymptotic behavior for $n$ large, and actually the full asymptotic expansion of the series,
depends only on the Taylor expansion of $f$ and $g$ at $x_0$.
This is the important principle of locality or localization which holds also for the
integrals that can be evaluated via the Laplace's method \cite{deburijn81}, \cite{bender99}, \cite{peter06}.

Following this principle, the main strategy of our proofs is the comparison of the series $I(n,\alpha)$ with a standard one whose summands
model the local behavior of $e^{-nf(k/n^{\alpha)}}g(k/n^{\alpha})$ around the global minimum point. In particular,
if the global minimum is attained at the boundary of the interval then, generically, the local model of $f$ is the linear function
$f(x)=\gamma x$ for some $\gamma>0$, while
if the global minimum point is $x_0 \neq 0$ then the local model of $f$ is generically the quadratic function $f(x)=\gamma (x-x_0)^2, \gamma>0$. 

In the first case, the standard series is simply the exponential sum
\begin{equation}\label{eq:exponentialsum}
E(n,\alpha,\gamma)=n^{1-\alpha} \sum_{k\geq0}e^{-\gamma n^{1-\alpha} k}=\frac{n^{1-\alpha}}{1-e^{-\gamma n^{1-\alpha}}} \, , \; \alpha,\gamma >0
\end{equation}the standard series is the Gaussian sum \footnote{In this case we let the summation variable $k$ run on all $\bb{Z}$ instead
of considering $k\geq0$. The difference between the two choices is immaterial
since the asymptotic behavior depends only on local contribution around $k \sim n^{\alpha}x_0$.}
\begin{equation}\label{eq:gaussian}
  Q(n,\alpha,\gamma, x_0)=n^{1-\alpha} \sum_{k \in \bb{Z}}e^{-n^{1-2\alpha} \gamma (k -n^{\alpha}x_0)^2}\, , \; \alpha,\gamma,x_0 >0 \; .
\end{equation}
 In fact, we show that the series we consider can be reduced, up to an exponentially small
relative error, to (\ref{eq:exponentialsum},\ref{eq:gaussian}) for an appropriate choice of the parameters.
We begin therefore the Section by computing the asymptotic behaviors of the Gaussian series
for all values of $\alpha$ (exponential sums being trivial).
Notice that all these Gaussian series can be written in terms of well-known Jacobi $\Theta$ functions \cite{bateman2,mumford83}, namely
\begin{equation}\label{eq:Qastheta}
 Q(n,\alpha,\gamma, x_0)=\sqrt{\frac{n \pi}{\gamma x_0} } \vartheta _3\left(-n^{\alpha} \pi ,e^{-\frac{n^{2 \alpha-1} \pi ^2}{\gamma x_0}}\right) \; .
\end{equation}
However, the asymptotic regimes we explore here
fall in general outside the ones normally considered because, for $\alpha \neq \frac{1}{2}$,
both the nome and the variable of the $\Theta$ functions depend on $n$.

\begin{lem}\label{lem:gaussian}
 The series $Q(n,\alpha,\gamma, x_0)$ (\ref{eq:gaussian}) has the following asymptotic behavior depending on $\alpha>0$:
 
 If $\alpha>\frac{1}{2}$ then
 \begin{equation}\label{eq:gaussianfirst}
  Q(n,\alpha,\gamma, x_0)=\sqrt{\frac{ \pi n}{\gamma}} (1+ O( e^{-\frac{2 \pi}{\gamma}n^{-1+2\alpha}})) \; .
 \end{equation}

If $\alpha=\frac{1}{2}$, then the Gaussian sum is a Jacobi theta function of fixed nome $q=e^{-\frac{\pi^2}{\gamma}}$
\begin{equation}\label{eq:gaussian1}
  Q(n,\frac{1}{2},\gamma, x_0)=\sqrt{\frac{\pi n}{\gamma x_0}}\theta_3(-\sqrt{n} \pi  x_0,e^{-\frac{\pi^2}{\gamma}}) \; .
\end{equation}
Notice that it is a strictly positive
periodic function of $(\sqrt{n}\pi x_0 -\lfloor \sqrt{n}\pi x_0 \rfloor)$.

If $\alpha<\frac{1}{2}$,
\begin{align}\label{eq:gaussiansmall}
  & Q(n,\alpha,\gamma, x_0)=n^{1-\alpha}\left( P(n,\alpha,\gamma,x_0) +
  O(e^{-  \gamma n^{1-2\alpha}}) \right) \;  \\ \label{eq:triangularwave}
 & P(n,\alpha,\gamma,x_0)=e^{-\gamma n^{1-2\alpha}t^2(n^{\alpha}x_0)}+e^{-\gamma n^{1-2\alpha}\big(1-t(n^{\alpha}x_0)\big)^2} \; ,
\end{align}
where $t(x)$ is the positive triangular wave of height $\frac{1}{2}$ and period $1$, namely
\begin{equation}\label{eq:t(x)}
 t(x)=\min\lbrace x-\lfloor x \rfloor \rbrace \; .
\end{equation}

 \begin{proof}
 We split the proof into to three cases, corresponding to $1-2\alpha$ negative, zero, or positive.
 The three must be dealt with using different techniques.
 
 In the case $1-2\alpha<0 $, the summand does not decrease fast uniformly on $n$. We transform the series into a tractable one
by means of the Poisson summation formula. In this special case, the Poisson summation formula is equivalent
to the following renowned (although slightly disguised) identity \cite{mumford83}
 \begin{equation*}
  \sum_{k \in \bb{Z}}e^{-n^{1-2\alpha} \gamma (k -n^{\alpha}x_0)^2}=n^{\alpha-1}
  \sqrt{\frac{ \pi n}{\gamma}}\sum_{q \in 2 \pi \bb{Z}}e^{-\frac{n^{-1+2\alpha}}{2\gamma }q(q+\frac{2 i x_0 \gamma}{n^{\alpha}})}
 \end{equation*}
 The term $q=0$ is equal to $n^{\alpha-1}\sqrt{\frac{ \pi n}{\gamma}}$. The total contribution of terms $q \neq 0$
 is easily bounded as
$$|\sum_{q \in 2 \pi \bb{Z}, q \neq 0} e^{-\frac{n^{-1+2\alpha}}{2\gamma }q(q+\frac{2 i x_0 \gamma}{n^{\alpha}})}|
\leq 2\sum_{q \in \bb{N} \setminus \lbrace 0 \rbrace}e^{-\frac{2 \pi n^{-1+2\alpha}}{ \gamma }q} =
O( e^{-\frac{n^{-1+2\alpha}2 \pi}{\gamma}}) \, .$$
 In the first inequality stems from the fact that $q^2 \geq q$ for $q\geq1$.
 
 The case $\alpha=\frac{1}{2}$ follows simply from the equation (\ref{eq:Qastheta}).
 
 We finally analyze the case $1-2\alpha>0$. The maximum of the summand is achieved when $n^{\alpha}x_0$ is closest to $0$, and the corresponding
 term contributes as $e^{-n^{1-2\alpha}t^2(n^{\alpha}x_0)} $. The second highest contribution is $ e^{-n^{1-2\alpha}\big(1-t(n^{\alpha}x_0)\big)^2}$.
 If $ t(n^{\alpha}x_0) \approx \frac{1}{2}$ then this second contribution cannot be neglected.
However, all other contributions are easily bounded, as
 \begin{align*}
  \sum_{k \in \bb{Z}}e^{-n^{1-2\alpha} \gamma (k -n^{\alpha}x_0)^2}- e^{-n^{1-2\alpha}\gamma t^2(n^{\alpha}x_0)} -
  e^{-n^{1-2\alpha}\gamma \big(1-t(n^{\alpha}x_0)\big)^2} \leq 
  2 \sum_{k \geq 1 }e^{-\gamma n^{1-2\alpha}k} =O(e^{-\gamma n^{1-2\alpha}}) \; .
 \end{align*}

\end{proof}
\end{lem}

Having collected the asymptotic behavior of the standard series we need, we can now state and prove our Theorems on
the Laplace's method for series $I(n,\alpha)$ (\ref{eq:INalpha}) for admissible pairs of functions $(f,g)$.

We start by considering the case of $f$ having a global minimum point belonging to the open interval $]0,\infty[$. 
In Theorem \ref{thm:laplacealpha}, we compute the leading asymptotic behavior of $I(n,\alpha)$ for a general $\alpha$.
In Theorem \ref{thm:laplace1}, we specialize to the case $\alpha=1$ and compute the full asymptotic expansion of $I(n,1)$, by
showing that it coincides at all orders with the well-known asymptotic expansion of the integral ${\cal{I}}(n)$.
We prove the latter Theorem by comparing the series and the integral by means of Euler-McLaurin formula.

\begin{thm}\label{thm:laplacealpha}
For $\alpha>0$ consider the series
\begin{equation*}
I(n,\alpha)=n^{1-\alpha}\sum_{k=0}^{\infty} e^{-n f(\frac{k}{n^{\alpha}})}g(\frac{k}{n^{\alpha}})
\end{equation*}
where $(f,g)$ is an admissible pair of functions. Assume furthermore that
\begin{itemize}
 \item[(i)] $f$ has a single global minimum point $x_0 \neq 0$,
 $f$ is three-times differentiable in a neighborhood of $x_0$ and $f''(x_0)>0$
 \item[(ii)] $\exists \e>0$
 such that $$\inf_{|x-x_0|>\e}f(x) > f(x_0) $$
 \item[(iii)]$g$ is differentiable in a neighborhood of $x_0$
\end{itemize}

Then the following asymptotic formula holds:
\begin{itemize}
 \item Case $\alpha>\frac{1}{2}$. For all $\beta \mbox{ s.t. } 0<\beta <\frac{1}{2}$,
 \begin{equation}\label{Inalpha}
I(n,\alpha)=\sqrt{\frac{2  \pi n}{f''(x_0)}}e^{-nf(x_0)}\big( g(x_0)+O(n^{-\beta}) \big) \,  \; .
\end{equation}
\item Case $\alpha=\frac{1}{2}$. For all $\beta \mbox{ s.t. } 0<\beta <\frac{1}{2}$,
\begin{align} \label{Inunmezzo}
 I(n,\frac{1}{2}) = e^{-nf(x_0)}
 \sqrt{\frac{2\pi n}{f''(x_0)}} \theta_3\left( -\sqrt{n} \pi  x_0,e^{-\frac{2\pi^2}{f''(x_0)}} \right)
 \big( g(x_0)+O(n^{-\beta}) \big) \, .
\end{align}
where $\theta_3$ is the third Jacobi Theta function.
\item Case $\frac{1}{3}<\alpha<\frac{1}{2}$.  For all $\beta \mbox{ s.t. } 0<\beta <\alpha$,
\begin{equation}\label{eq:INsmallaplha} 
 I(n,\alpha)=n^{1-\alpha}e^{-n f(x_0)} P\left( n,\alpha,\frac{f''(x_0)}{2} ,x_0 \right) 
 \big( g(x_0)+O(n^{1-2\alpha-\beta}) \big)\; , 
\end{equation}
where $P(n,\alpha,\gamma,x_0)$ is the oscillating sequence defined in equation (\ref{eq:triangularwave}) above.
\item Case $0<\alpha\leq \frac{1}{3}$. A weaker statement holds
\begin{equation}
 \lim_{n \to \infty}e^{nf(x_0)}n^{\alpha-1}\big(I(n,\alpha)-P(n,\alpha,\frac{f''(x_0)}{2},x_0) \big) =0 \; .
\end{equation}
\end{itemize}

\begin{proof}
If we multiply the series by $\mbox{sign} (g(x_0))e^{nf(x_0)}$ we reduce to the case $f(x_0)=0$ and $g(x_0)\geq0$ which we assume to hold.
The hypothesis that $(f,g)$ are admissible implies that there exist $n\geq 1,C,M>0$ such that $|I(n,\alpha)|\leq C n^{M}$ for all $n$ big enough.

The strategy of the proof goes as follows.
First I) we show that terms not sufficiently close to the maximum are exponentially suppressed, then II) we show can be approximated by
the Gaussian sum $Q(n,\alpha,\frac{f''(x_0)}{2},x_0)$ up to the stated error.

I) 
After hypotheses (i,ii),
we can choose a $\mu$, $0<\mu< f''(x_0)$ and $\delta>0$, such that for any
$\delta\leq \delta_0$ then
$f(x) \geq \mu \delta^2$ for all  $|x-x_0| \geq \delta$.

Therefore if $|k-n^{\alpha}x_0|\geq n^{\alpha} \delta$ then
$$
n f(\frac{k}{n^{\alpha}}) \geq (n-m_0) \mu \delta^2 + m_0 f(\frac{k}{n^{\alpha}}) \, , \forall m_0 \in \bb{R} \; .
$$
Let $m_0$ big enough such that estimate (\ref{eq:boundadmissible}) holds, then
\begin{equation}\label{eq:boundtheorem1}
|\sum_{|\frac k{n^{\alpha}}-x_0|\geq \delta} e^{-n f(\frac{k}{n^{\alpha}})}g(\frac{k}{n^{\alpha}}) |\leq
e^{-(n-m_0)\mu \delta^2} \sum_{k\geq 0}
| e^{-m_0 f(\frac{k}{n^{\alpha}})}g(\frac{k}{n^{\alpha}})|\leq
C' n^{M'} e^{-\mu n \delta^2} \; ,
\end{equation}
for some $C',M' >0$. Here we used estimate (\ref{eq:boundadmissible}) to deduce the last inequality.

A slightly stronger version of bound (\ref{eq:boundtheorem1}) holds for the Gaussian series  $Q(n,\alpha,\gamma,x_0), \gamma>0$.
In fact,
\begin{equation}\label{eq:bound2theorem1}
 |Q(n,\alpha,\gamma,x_0)-n^{1-\alpha}\sum_{ |k-n^{\alpha}x_0|\geq n^{\alpha} \delta }e^{-\gamma n^{1-2\alpha}(k-n^{\alpha}x_0)^2}|\leq 2
 \frac{n^{1-\alpha}e^{-\gamma n \delta^2}}{1-e^{-\gamma n^{1-2\alpha} }} \leq C' n e^{-\gamma n \delta^2} \; .
\end{equation}

Inequalities (\ref{eq:boundtheorem1},\ref{eq:bound2theorem1}) tell us that
the total contribution of all terms away from $n^{\alpha}x_0$ is exponentially small as $n\to +\infty$ if $\delta$ is
any small enough fixed positive number. This is also the case if we let $\delta$ depend on $n$
according to the law $\delta = n^{-\beta} $ with $0<\beta<\frac{1}{2}$. This latter choice will be convenient later on in the proof.

II)We now give a bound for each term in the sum when $k$ is close to $n^{\alpha}x_0$ using
a simple Taylor expansion. Here, to avoid a cumbersome notation, we suppose $g(x_0)\neq 0$. The few modifications needed to
consider the case $g(x_0)=0$ are trivial.

Since $g$ is differentiable at $x_0$ and $f$ has three derivatives at $x_0$ then, for every $y$ small enough, there exists a $\kappa>0$ such that
if $|x-x_0|\leq y$ then
$$
|f(x)-f(x_0)+(x-x_0)^2 f''(x_0) | \leq \kappa f''(x_0) (x-x_0)^2 y \mbox{ and } |g(x)-g(x_0)|\leq \kappa y \; .
$$
Therefore if $|k-n^{\alpha}x_0|\leq \delta n^{\alpha} \leq y n^{\alpha} $
\begin{align*}
& e^{-n f(\frac{k}{n^{\alpha}})}g(\frac{k}{n^{\alpha}}) \leq e^{-n^{1-2\alpha}
 \frac{f''(x_0)}{2}(1-\kappa\delta)(k-n^{\alpha})^2}(g(x_0)+\kappa \delta) \\
 & e^{-n f(\frac{k}{n^{\alpha}})}g(\frac{k}{n^{\alpha}}) \leq e^{-n^{1-2\alpha}
 \frac{f''(x_0)}{2}(1+\kappa\delta)(k-n^{\alpha})^2}(g(x_0)-\kappa \delta)
\end{align*}
Let us analyze the upper-bound.
Letting $\delta=n^{-\beta}$ $\beta>0$, we have that for any  $n$ is big enough then 
\begin{equation}\label{eq:thm1truncatedbound}
 \sum_{|k-n^{\alpha}x_0|\leq n^{\alpha-\beta}} e^{-n f(\frac{k}{n^{\alpha}})}g(\frac{k}{n^{\alpha}})\leq
 \sum_{|k-n^{\alpha}x_0|\leq n^{\alpha-\beta}}
 e^{-n^{1-2\alpha} (k-n^{\alpha}x_0)^2 \frac{f''(x_0)}{2}(1- \kappa n^{-\beta} )}\big( g(x_0)+\kappa n^{-\beta}\big)
\end{equation}
Using (\ref{eq:boundtheorem1},\ref{eq:bound2theorem1}) we obtain
\begin{eqnarray}\label{eq:thm1semiboundp}
 I(n,\alpha)\leq Q\left( n,\alpha,\frac{f''(x_0)}{2}\big(1- c n^{-\beta}\big),x_0 \right) \big( g(x_0)+c n^{-\beta} \big)+
 O(n^{M'} e^{-\mu n^{1-2\beta}}) \mbox{ for some } M'>0 \; .
\end{eqnarray}
Here we have merged the the error terms coming from (\ref{eq:boundtheorem1},\ref{eq:bound2theorem1}) in a single one,
considering that $0<\mu \leq f''(x_0)/2 $. Notice that the bound is effective only for any $\beta, 0<\beta<\frac{1}{2}$, a constraint
we assume to hold.

By the same reasoning, for the lower bound we obtain
\begin{eqnarray}\label{eq:thm1semiboundm}
 I(n,\alpha)\geq Q\left( n,\alpha,\frac{f''(x_0)}{2}\big(1+ \kappa n^{-\beta}\big),x_0 \right)
 \big( g(x_0)-\kappa n^{-\beta} \big)+
O(n^{M'} e^{-\mu n^{1-2\beta}}) \mbox{ for some } M'>0 \;  .
\end{eqnarray}

We now complete our analysis by considering the four cases $\alpha>1/2,\alpha=1/2,1/3<\alpha<1/2,\alpha\leq 1/3$ separately.

In the case $\alpha>1/2$, using (\ref{eq:gaussianfirst}) we have 
$$
Q\left( n,\alpha,\frac{f''(x_0)}{2}\big(1 \pm \kappa n^{-\beta}\big),x_0 \right) \big( g(x_0)\pm \kappa n^{-\beta}) \big)=
\sqrt{\frac{2 \pi n}{f''(x_0)}} \big( g(x_0)+O (n^{-\beta}) \big) \; .
$$
The latter estimate together with inequalities (\ref{eq:thm1semiboundp},\ref{eq:thm1semiboundm}) implies the thesis.

Similarly in the case $\alpha=\frac{1}{2}$, using (\ref{eq:gaussian1}) we have
\begin{eqnarray*}
Q\left( n,1/2,\frac{f''(x_0)}{2}\big(1 \pm \kappa n^{-\beta}\big),x_0 \right) \big( g(x_0)\pm \kappa n^{-\beta} \big) =
\sqrt{\frac{2\pi n}{f''(x_0)}}\theta_3(-\sqrt{n} \pi  x_0,e^{-\frac{2\pi^2}{f''(x_0)}})\big( g(x_0)+O(n^{-\beta}) \big) ,
\end{eqnarray*}
which, together with inequalities (\ref{eq:thm1semiboundp},\ref{eq:thm1semiboundm}), proves the thesis.
The latter inequality stems from the fact that $\theta_3$ and its derivatives are bounded (in fact periodic) function of the first
variable, namely
$$
\lim_{\e \to 0} \sup_{x \in \bb{R}} |\partial_y \theta_3(x,y_0 + \e)| = \sup_{x \in [0,1]}  |\partial_y \theta_3(x,y_0 + \e)| <\infty \; .
$$

For $\alpha <\frac{1}{2}$ the Gaussian series $Q(n,\alpha,\gamma,x_0)$ oscillates between $n^{1-\alpha}\left(1+O(e^{-\gamma n^{1-2\alpha}})\right)$
and exponentially small values.
More precisely, from
(\ref{eq:gaussiansmall}) we have
\begin{eqnarray*}
P\left(n,\alpha,\frac{f''(x_0)}{2}\big(1\pm \kappa n^{-\beta}\big),x_0\right) \geq \exp \lbrace- n^{1-2\alpha}
\frac{f''(x_0)}{4} (1\pm \kappa n^{-\beta}) \rbrace \; .
\end{eqnarray*}
In order to compare $I(n,\alpha)$ with the Gaussian series we needs to check that
the error term in inequalities (\ref{eq:thm1semiboundp},\ref{eq:thm1semiboundm}) are negligible with respect to it.
To this aim we need to assume that $\beta <\alpha$.

From the definition of the function $P$, equation (\ref{eq:gaussiansmall}), it follows that for any $\beta>0$
\begin{eqnarray*}
P\left( n,\alpha,\frac{f''(x_0)}{2}\big(1 \pm \kappa n^{-\beta}\big),x_0 \right) \big( g(x_0)\pm \kappa n^{-\beta} \big) =\\
 P\left( n,\alpha,\frac{f''(x_0)}{2},x_0 \right) \big( g(x_0)+O(n^{1-2\alpha -\beta}) \big) \; .
\end{eqnarray*}
If $1/3<\alpha<1/2$ then we can find $\beta<\alpha$ such that $n^{1-2\alpha -\beta} \to 0$. The thesis is then proved.

However, if $0<\alpha \leq 1/3$ then  $n^{1-2\alpha -\beta} \nrightarrow 0$ for any $\beta<\alpha$. To prove the thesis in this case,
it is sufficient to show that for any $c>0$
\begin{equation}\label{eq:thm1last}
 \lim_{n \to \infty}n^{\alpha-1}\big(P(n,\alpha,\frac{f''(x_0)}{2}\pm \kappa n^{-\beta},x_0)-P(n,\alpha,\frac{f''(x_0)}{2},x_0) \big) =0
\end{equation}
To this aim we fix an $\e , \, 0<\e<\alpha$ and for any sequence $n\to \infty$, we define $n_l$ to be the subsequence such that
$t^2(n_l^{\alpha}x_0) \geq  n_l^{2\alpha-1+\e} $ and $n_m$ to be the complementary subsequence.
We prove that for both subsequences above limit holds.

From (\ref{eq:triangularwave}), it follows directly that for any $c\in \bb{R}$
 $$
 n_l^{\alpha-1}\big(P(n,\alpha,\frac{f''(x_0)}{2}\pm \kappa n_l^{-\beta},x_0))
 \leq 2 e^{-n_l^{\e}(\frac{f''(x_0)}{2}\pm \kappa n_l^{-\beta} )} \; .
 $$
Therefore both terms in (\ref{eq:thm1last}) tend to zero, so their difference.

Conversely, for $ m \to \infty$ and any $\beta >\e$ we have that 
$$
\frac{P(n_m,\alpha,\frac{f''(x_0)}{2}\pm \kappa n_m^{-\beta},x_0)}{P(n_m,\alpha,\frac{f''(x_0)}{2},x_0)}=1+O(n_m^{\e-\beta}) \; .
$$
Since $P(n_m,\alpha,\frac{f''(x_0)}{2},x_0))$ is bounded then (\ref{eq:thm1last}) holds.
\end{proof}

\end{thm}


\begin{cor}\label{cor:laplaceslapha}
With the same hypothesis as in Theorem \ref{thm:laplacealpha}. Let $f,g$ be measurable functions for
which ${\cal{I}}(n)$ converges absolutely for $n$ big enough.
If $\alpha>\frac{1}{2}$ and $g(x_0) \neq 0$ then
$I(n,\alpha) \sim {\cal{I}}(n) \equiv  n \int_0^{\infty} e^{-nf(x)}g(x)$.
\begin{proof}
 The thesis follows from the standard Laplace's method 
 for integrals \cite{peter06}, which states that ${\cal{I}}(n)\sim \sqrt{\frac{2\pi n}{f''(x_0)}}g(x_0)$. 
\end{proof}

\end{cor}

\begin{cor}\label{cor:fullasymptotic}
 Let $f,g$ satisfy all hypotheses of Theorem \ref{thm:laplacealpha} and, in addition to that, let us assume
 that $g(x_0)\neq 0$. Then 
\begin{eqnarray*}
I(n,\alpha) \sim e^{-nf(x_0)} g(x_0)
\left\lbrace 
\begin{matrix}
 \sqrt{\frac{2 \pi n}{f''(x_0)} } \, , \; \mbox{ if } \alpha >\frac{1}{2} \\
 \sqrt{\frac{2\pi n}{f''(x_0)}}\theta_3(-\sqrt{n} \pi  x_0,e^{-\frac{2\pi^2}{f''(x_0)}}))\, , \; \mbox{ if } \alpha=\frac{1}{2}\\
 n^{1-\alpha} P(n,\alpha,\frac{f''(x_0)}{2},x_0) \, , \; \mbox{ if } \frac{1}{3}<\alpha<\frac{1}{2} 
\end{matrix} \right. \, .
\end{eqnarray*}
\begin{proof}
The Corollary is a direct consequence of Theorem \ref{thm:laplacealpha}.
\end{proof}

 \end{cor}

In Theorem \ref{thm:laplace1} below we assess the complete asymptotic expansion of $I(n,1)$ by directly comparing the series
$I(n,1)$ with the integral ${\cal{I}}(n)$ by means of
Euler-Mc Laurin summation formula. To this aim, we needs some preparatory results, Lemma \ref{lem:expansionderivative} and
Proposition \ref{prop:laplacezeroes}.

\begin{lem}\label{lem:expansionderivative}
 Let $f,g$ be two functions of one variable with $k$ continuous derivatives then
 \begin{equation}
  \frac{\partial^k}{\partial y^k} e^{- n f(y/n)}g(y/n)=\left( \sum_{l=0}^k n^{-l} (f'(y/n))^{\ma{k-2l}{0}}Q_{k,l}(y/n) \right) e^{- n f(y/n)}
 \end{equation}
 where $\ma j0=\max \lbrace j, 0 \rbrace$ and  $Q_{k,l}(x)\equiv Q_{k,l}(f'(x),\dots f^{(k)}(x), g(x) ,\dots g^{(k)}(x)) $
 are differential polynomials.
 \begin{proof}
 We notice that the thesis is equivalent to
 \begin{equation*}
  \frac{\partial^k}{\partial y^k} e^{- n f(y)}g(y)=\left(\sum_{l=0}^k n^{k-l} (f'(y))^{\ma{k-2l}0}Q_{k,l}(y)\right) e^{- n f(y)}
 \end{equation*}
  We prove the latter statement by induction.
  
  If $k=1$ the Lemma is trivially  true. Suppose the Lemma holds for $k$ we show that it holds for $k+1$. In fact
  \begin{align*}
 &  \frac{\partial^{k+1}k}{\partial y^{k+1}} e^{- n f(y)}g(y)= \frac{\partial}{\partial y}\frac{\partial^k}{\partial y^k} e^{- n f(y)}g(y) =
 \left( \sum_{l=0}^k n^{k-l+1} (f'(y))^{\ma{k-2l}0+1}Q_{k,l}(y)+ \right. \\
  &  \left. 
  \sum_{l=0}^kn^{k-l} (f'(y))^{\ma{k-2l}0-1} 
   \big( \ma{k-2l}0 f''(y)  Q_{k,l}(y) +    f'(y) \partial_y Q_{k,l}(y) \big) \right) e^{-n f(y)}  = \\
&  \left( \sum_{l=0}^{k+1} n^{k+1-l} (f'(y))^{\ma{k+1-2l}0} Q_{k+1,l}(y) \right) e^{-n f(y)} \; .
  \end{align*}
Here $$Q_{k+1,l}=\big(f'(y)\big)^{(k-2l,0)}Q_{k,l}+\ma{k-2l+2}0 f''(y)  Q_{k,l-1}(y) + f'(y) \partial_y Q_{k,l-1}(y) \; ,$$
where $Q_{k,k+1}=Q_{k,-1}=0$ and $(k-2l,0)=1 \mbox{ if } k-2l < 0, 0 \mbox{ otherwise}$.
Therefore $Q_{k+1,l}$ is a differential polynomial in the variables $f',\dots f^{(k+1)}$,
$g ,\dots, g^{(k+1)} $
and the thesis is proven.
 \end{proof}
\end{lem}
The last piece of mathematical technology we need to prove Theorem \ref{thm:laplace1} is a natural extension of the Laplace's method.
\begin{pro}\label{prop:laplacezeroes}
Let $f,g$ are measurable functions satisfying the same hypotheses of Theorem \ref{thm:laplacealpha}, and in addition to them,
\begin{equation}
 g(x)=(x-x_0)^k r(x)
\end{equation}
for some $k\geq0$ and some continuous function $r(x)$.

Then for any $x^*>x_0$
 \begin{equation*}
n\int_0^{x^*}e^{-nf(x)} g(x) dx = O(n^{\frac{1-l}{2}})  \; .
 \end{equation*}
 
 \begin{proof}
  See e.g.  \cite{erdelyi56} page 37.
 \end{proof}

\end{pro}

\begin{thm}\label{thm:laplace1}
With the same hypothesis of Theorem \ref{thm:laplacealpha}. Consider the series
\begin{equation*}
I(n)\equiv I(n,1)= \sum_{k=0}^{\infty}e^{-n f(\frac{k}{n})}g(\frac{k}{n})
\end{equation*}
where $f,g$ are measurable functions such that for some $m \geq 0$
\begin{itemize}
 \item[(i)]$f$ has $m+2$ continuous derivatives in a neighborhood of $x_0$.
 \item[(ii)]$g$ has $m+1$ continuous derivatives in a neighborhood of $x_0$.
\end{itemize}
For any $x^*>x_0$
\begin{equation}\label{Incexpansion}
e^{n f(x_0)}\big( I(n,\alpha)-n \int_0^{x^*}e^{-nf(x)} g(x) dx \big)=o(n^{-\lfloor \frac{m}{2} \rfloor+\frac{1}{2}})
\end{equation}
Due to the well-known \cite{peter06} asymptotic expansion of ${\cal{I}}(n)$, formula (\ref{Incexpansion}) implies that
 \begin{equation}\label{thmasymptotics}
I(n)=\sqrt{\frac{2 n \pi}{f''(x_0)}}e^{-nf(x_0)}\left( g(x_0)+\sum_{l=1}^{\lfloor \frac{m}{2} \rfloor } a_l n^{-l}+o (n^{-m}) \right)
\end{equation}
where $a_l,l=1,\dots,m$ depends on the first $l+2,l$ derivatives of $f,g$ at $x_0$. 

\begin{proof}
 
If we multiply the series by $e^{nf(x_0)} $ we reduce to the case $f(x_0)=0$.

In the proof of Theorem \ref{thm:laplacealpha} we proved that
the total contributions of terms where $|k -nx_0|> n \delta $ for a fixed $\delta>0$ $o(n^{-K})$ for all $K>0$.
The same principle holds for the integral \cite{peter06} : for any $x_1,x_2 >x_0$
$$
\int_0^{x_1}e^{-nf(x)} g(x) dx=\int_0^{x_2}e^{-nf(x)} g(x) dx + o(n^{-K}) \, , \; \forall K>0 \; .
$$
Therefore to prove the thesis it will suffice to show that
\begin{equation}\label{eq:thm2tobeshown}
 \sum_{k=0}^{n n_0}e^{-n f(k/n)}g(k/n)=\int_0^{n_0}e^{-nf(x)} g(x) dx + o(n^{-\lfloor \frac{m}{2} \rfloor+\frac{1}{2}}) \; .
\end{equation}
Moreover, without losing generality, we can assume that $f,g$ have $m+2,m+1$ continuous derivative in the whole segment $[0,n_0]$.
 
To prove the above statement, we use Euler(-McLaurin) summation formula, according to which (see e.g. \cite{apostol99}) for any (integrable) function $h$
with $2p$ (resp. $2p+1$) continuous derivatives the following formula holds
\begin{align}\label{eq:eulermclaurin}
 & \sum_{k=0}^Nh(k)=\int_0^n h(y)dy+\frac{h(N)-h(0)}{2}-\sum_{k=1}^p \frac{B_{2k}}{(2k)!}(h^{(2k-1)}(N)-h^{(2k-1)}(0))+R\\ \nonumber
 & R = C_{2p} \int_0^N B_{2p}(y)h^{(2p)}(y) dy  = C_{2p+1} \int_0^N B_{2p+1}(y)h^{(2p+1)}(y) dy \, .  
\end{align}
Here $B_{2k}$ is the 2k-th Bernoulli number,
$B_{j}(y)$ is the j-th periodic Bernoulli function, which is a $j-2$ times differentiable
periodic function of period $1$ and finally $C_j$ is a universal constant; see \cite{apostol99} for the precise definitions.

We apply the formula to $h(y)=e^{-n f(y/n)}g(y/n)$ and $N = n n_0$ to get
$$
I(n)=n\int_0^{n_0}e^{-nf(x)}g(x) dx+ \, \mbox{ remainder }
$$
where the remainder is made of contribution from the evaluation at the endpoint of the interval -which are discarded because exponentially small-
and the term $R$, involving the integral of $h^{(m+1)}$ as by hypothesis $h$ has $m+1$ continuous derivatives.

In order to show that $R=o(n^{-\frac{m-1}{2}})$,
we first make use of Lemma \ref{lem:expansionderivative} to obtain a first expansion of $R$ in power series of $n$
\begin{align*}
 R=n\int_0^{n_0} B_{m+1}(xn) e^{-n f(x)} \left( \sum_{l=0}^{m+1} n^{-l} (f'(x))^{\ma{m+1-2l}0}Q_{m+1,l}(x) \right)  dx \; ,
\end{align*}
where $\ma j0=\max \lbrace j, 0 \rbrace$.

Since $f'(x_0)=0$ ($x_0$ is a minimum for $f$), then $(f'(x))^{\ma{m+1-2l}0}Q_{l,k}(f(x),g(x))=(x-x_0)^{\ma{m+1-2l}0}r_{m+1,l}(x)$
for some continuous function $r_l$.

After Proposition \ref{prop:laplacezeroes}, we can conclude that there are $C_0,\dots,C_{m+1}>0$ such that
\begin{align*}
 |R|\leq n^{1/2} \left( \sum_{l=0}^{m+1} C_l n^{-l} n^{-\ma{m+1-2l}0/2} \right) = O(n^{-m/2}) = o(n^{-\frac{m-1}{2}})\; .
\end{align*}
 
\end{proof}

\end{thm}

\begin{cor}
If $f,g$ satisfy the hypotheses of Theorem \ref{thm:laplace1} and in addition they are smooth
in the neighborhood of the unique  global minimum point $x_0$ of $f$,
then $I(n)$ admits an asymptotic expansion in odd (negative) powers of $n^{\frac{1}{2}}$ that coincides with the asymptotic expansion
of ${\cal{I}}(n)$.
\end{cor}

We turn now our attention to the asymptotic evaluation of $I(n,\alpha)$ when the global minimum point is $0$.
In Theorem \ref{thm:watsonalpha} we deal with the leading asymptotic behavior of $I(n,\alpha)$ for general
$\alpha$, and in Theorem \ref{thm:Watson1} we compute
the full asymptotic expansion in the case $\alpha=1$.

\begin{thm}\label{thm:watsonalpha}
For $\alpha>0$ consider the series
\begin{equation*}
I(n,\alpha)=n^{1-\alpha}\sum_{k=0}^{\infty} e^{-n f(\frac{k}{n^{\alpha}})}g(\frac{k}{n^{\alpha}})
\end{equation*}
where $(f,g)$ is an admissible pair of functions. Assume furthermore that
\begin{itemize}
 \item[(i)]$f$ has a single global minimum at $x=0$, it is differentiable in a neighborhood of $0$ and $f'(0)>0$
 \item[(ii)]there exists an $\e$ such that $$\inf_{x>\e}f(x) > f(0)$$
 \item[(iii)]$g$ is differentiable in a neighborhood of $0$
\end{itemize}
Then the following asymptotic formula holds:
\begin{itemize}
 \item Case $\alpha>1$. For all $\beta \mbox{ s.t. } 0<\beta <1 \mbox{ et } \beta<\alpha-1$,
\begin{equation}\label{Jnalpha}
I(n,\alpha)=\frac{e^{-nf(0)}}{f'(0)}(g(0)+O(n^{-\beta}))  \;.
\end{equation}
\item Case $\alpha=1$.
For all $ \beta \mbox{ s.t. } 0< \beta < 1$
\begin{equation}\label{Jnuno}
  I(n) = \frac{e^{-n f(0)}}{1-e^{-f'(0)}} (g(0)+O(n^{-\beta}))  \; .
\end{equation}
\item Case $0<\alpha<1$. For all $\beta \mbox{ s.t. } 0< \beta < 1$
\begin{equation}\label{Jnmuno}
 I(n,\alpha)=n^{\alpha-1}e^{-n f(0)}g(0)  (1 + O(n^{-\beta})) \; .
\end{equation}
\end{itemize}
\begin{proof}
The proof follows the same strategy as the proof of Theorem \ref{thm:laplacealpha}.

If we multiply the series by $e^{nf(x_0)} \mbox{sign}(g(0))$ we reduce to the case $f(0)=0, g(0)\geq 0$.
Since $(f,g)$ are admissible, $\exists n_0,C, M >0$ such that for $n \geq n_0$, $I(n,\alpha)\leq C n^{M}$.

We notice that only contributions localized at $k=0$ do matter asymptotically. 
Indeed,
there exist $\mu,\delta_0,C',M'$ such that for any $\delta \leq \delta_0$
\begin{equation}\label{eq:boundtheorem3}
|\sum_{k\geq n^{\alpha} \delta} e^{-n f(\frac{k}{n^{\alpha}})}g(\frac{k}{n^{\alpha}}) |\leq   C' n^{M'} e^{-\mu n \delta} \; .  
\end{equation}
To prove the latter bound we follow the same steps as in Theorem \ref{thm:laplacealpha}. Namely we notice, after hypotheses (i,ii) on $f$,
that if $\delta>0$ is small enough there exists a $0<\mu< f'(0)$ such that
$f(x) \geq \mu \delta$ for all  $x \geq \delta$.
Therefore for any $\delta$ sufficiently small
$$
n f(\frac{k}{n^{\alpha}}) \geq (n-m_0) \mu \delta + m_0 f(\frac{k}{n^{\alpha}}) ; .
$$
Estimate (\ref{eq:boundtheorem3}) follows from the latter bound, by choosing $m_0$ big enough so that bound (\ref{eq:boundadmissible}) holds.

If $\delta$ is fixed or $\delta = n^{-\beta} $ with $0<\beta<1$, then the total contribution of all terms for $k \geq n^{\alpha}\delta$
is exponentially small.
The same principle holds also for the exponential series $E(n,\alpha,\gamma)$  (\ref{eq:exponentialsum}) which is just a special case
of $I(n,\alpha)$ with $f(x)=\gamma x, g(x)=1$: we have the slightly stronger bound
\begin{equation}\label{eq:bound2theorem3}
|\sum_{k\geq n^{\alpha} \delta} e^{-\gamma n^{1-\alpha} k} | \leq C n^{M} e^{-\gamma n \delta} \, \; \mbox{ for some } C,M>0 \;. 
\end{equation}

As in the proof of Theorem \ref{thm:laplacealpha}, we compare, the truncated $I(n,\alpha)$ with the exponential series.
Again, to avoid a cumbersome notation, we suppose $g(0)\neq 0$. The few modifications needed to
consider the case $g(x_0)=0$ are trivial.
Since $f$ is twice differentiable at $0$ and $g$ is differentiable at $0$ then
then, for every $y$ small enough, there exists a $\kappa>0$ such that
if $x\leq y$ then
$$
|f(x)-(x-x_0) f'(0) | \leq \kappa f'(0) x y \mbox{ and } |g(x)-g(0)|\leq \kappa y \; .
$$
Therefore if $k \leq \delta n^{\alpha} \leq y n^{\alpha} $
\begin{align*}
& e^{-n f(\frac{k}{n^{\alpha}})}g(\frac{k}{n^{\alpha}}) \leq e^{-n^{1-\alpha}
 f'(0)(1-\kappa\delta)k}(g(0)+\kappa \delta) \\
 & e^{-n f(\frac{k}{n^{\alpha}})}g(\frac{k}{n^{\alpha}}) \geq e^{-n^{1-\alpha}
 f'(0)(1+\kappa\delta)k}(g(0)-\kappa \delta)
\end{align*}
Let us analyze the upper-bound.
Letting $\delta=n^{-\beta}$ $\beta>0$, we have that for any  $n$ is big enough 
\begin{equation}\label{eq:thm3truncatedbound}
 \sum_{k\leq n^{\alpha-\beta}} e^{-n f(\frac{k}{n^{\alpha}})}g(\frac{k}{n^{\alpha}})\leq \sum_{k\leq n^{\alpha-\beta}}
 e^{-n^{1-\alpha} ( f'(0)- \kappa n^{-\beta})k}(g(x_0)+\kappa n^{-\beta}))
\end{equation}
Using (\ref{eq:boundtheorem3},\ref{eq:bound2theorem3}) we obtain
\begin{eqnarray} \label{eq:thm3semiboundp}
 I(n,\alpha)\leq E\left( n,\alpha,f'(0)- \kappa n^{-\beta} \right) \big( g(0)+\kappa n^{-\beta} \big)+
 O(n^{M'} e^{-\mu n^{1-\beta}}) \mbox{ for some } M'>0 \; .
\end{eqnarray}
Here we have merged the error terms from (\ref{eq:boundtheorem3},\ref{eq:bound2theorem3}) into a single one, considering that $\mu <f'(0)$.
Notice that the bound is effective only for any $\beta, 0<\beta<1$, a constraint
we assume to hold.

The same analysis shows that we have the lower bound
\begin{eqnarray}\label{eq:thm3semiboundm}
 I(n,\alpha)\leq E\left( n,\alpha,f'(0)+ \kappa n^{-\beta} \right) \big( g(0)-\kappa n^{-\beta} \big)+\
 O(n^{M'} e^{-\mu n^{1-\beta}}) \mbox{ for some } M'>0 \; .
\end{eqnarray}

After equation (\ref{eq:exponentialsum}) for $E(n,\alpha,\gamma)$, the following formulas hold
\begin{eqnarray*}
 n^{\alpha-1}E\left( n,\alpha,f'(0)\pm \kappa n^{-\beta} \right) \big( g(0)\mp \kappa n^{-\beta} \big) 
   =
 \left\lbrace 
\begin{matrix}
  \frac{g(0) + O (n^{-\beta})}{f'(0)}   \, , \; \mbox{ if } \alpha>1 \\
\frac{g(0)+O( n^{-\beta})}{1-e^{-f'(0)}} \, , \; \mbox{ if } \alpha=1\\
n^{1-\alpha}g(0) ( 1+O(n^{-\beta}) )\, , \; \mbox{ if } \alpha<1 
\end{matrix} \right.
\end{eqnarray*}
The thesis follows after the latter formula and the estimate (\ref{eq:thm3semiboundp},\ref{eq:thm3semiboundm}).
 
\end{proof}

\end{thm}

Before introducing and proving Theorem \ref{thm:Watson1} about the asymptotic expansion of the series $ I(n,1)$,
we introduce here a last estimate we need to prove them. 
\begin{lem}\label{lem:prepthm3}
 Let $f,g:[0,\infty[ \to \bb{R}$ be two functions such that, for some $m \geq 0$,
 \begin{itemize}
  \item $f$ has $m+2$ continuous right derivatives at $x=0$ and $f(0)=0$
  \item $g$ has $m+1$ continuous right derivatives at $x=0$
 \end{itemize}
 Fix $\beta>\frac{1}{2}$. On the set $\lbrace n\geq 1 , k\leq n^{1-\beta} \rbrace$ the following estimate hold
 \begin{eqnarray*}
  \left| e^{-nf(k/n)}g(k/n)- e^{-f'(0)k} \big( g(0)+ \sum_{l=1}^m P_l(k) n^{-l} \big) \right|\leq  n^{-m-1} e^{-f'(0)k} R(k) \; .
 \end{eqnarray*}
where 
\begin{eqnarray*}
 P_l(k)=e^{f'(0)k}\frac{1}{l!}\left(\frac{\partial^l}{\partial (n^{-1})^l}e^{-nf(k/n)}g(k/n)\right)_{n=\infty} \; .
\end{eqnarray*}
and $R(k)$ is some polynomial in $k$.

\begin{proof}
 
Notice that $-nf(k/n)+f'(0)k=O(n^{1-2\beta}) $. Since $\beta >\frac{1}{2}$ then we can expand $e^{-nf(k/n)+f'(0)k}$,
in power series of $ (-nf(k/n)+f'(0)k)$. The thesis then follows by applying Taylor's formula to the function $F(n;k)=e^{-nf(k/n)+f'(0)k}g(k/n)$.
\end{proof}

\end{lem}

\begin{thm}\label{thm:Watson1}
 Consider the series
\begin{equation*}
I(n)=\sum_{k=0}^{\infty}e^{-n f(\frac{k}{n})}g(\frac{k}{n})
\end{equation*}
 where $f,g$ satisfy the same hypotheses as in Theorem \ref{thm:watsonalpha}, and in addition to them, there
exists an $m \geq 0$ such that
\begin{itemize}
 \item[(i)]$f$ has $m+2$ continuous derivatives in a (right-)neighborhood of $0$
 \item[(ii)]$g$ has $m+1$ continuous derivatives in a (right-)neighborhood of $0$
\end{itemize}
then
\begin{equation}\label{eq:watsonregularcomplete}
  I(n,1) = e^{-n f(0)} \left( \frac{g(0)}{1-e^{-f'(0)}}+\sum_{l=1}^m a_l n^{-l} + o(n^{-m}) \right)
 \end{equation}
 where
 \begin{eqnarray}\label{eq:alpha}
a_l= \sum_{k\geq0}e^{- k f'(0) }P_l(k)  , \, 
P_l(k)=\left(\frac{\partial^l}{\partial (n^{-1})^l}e^{-n\sum_{j=2}^{m+1}\frac{f^{(l)}(0)}{j!}\frac{k^j}{n^j}}g(k/n)\right)_{n=\infty} \;.
 \end{eqnarray}
\begin{proof}
As usual, we multiply the series by $e^{nf(x_0)} \mbox{sign}(g(0))$ we reduce to the case $f(0)=0, g(0)\geq 0$.
Since $(f,g)$ are admissible, $\exists n_0,C, M >0$ such that for $n \geq n_0$ $I(n,\alpha)\leq C n^{M}$.

A trivial extension of the estimate (\ref{eq:bound2theorem3}) implies that if
$\beta <1 $ then for any polynomial $P(k)$ there exist $C',M'$ such that
\begin{eqnarray*}
 |\sum_{k/n\geq n^{-\beta}}e^{-\gamma k}P(k)| \leq  C' n^{M'} e^{-\gamma n^{1-\beta}}\; .
\end{eqnarray*}

The thesis follows from the latter bound, estimate (\ref{eq:boundtheorem3}) and Lemma (\ref{lem:prepthm3}).
\end{proof}

\end{thm}

\begin{cor}\label{cor:watsonsalpha}
With the same hypothesis of Theorem \ref{thm:watsonalpha}.
Let $f,g$ be measurable functions, $\alpha>1$ and $g(x_0) \neq 0$ then
for any $x^*>0$
$$I(n,\alpha) \sim   n \int_0^{x^*} e^{-nf(x)}g(x)\;.$$
\begin{proof}
The standard Watson's Lemma \cite{peter06} implies that $$  n \int_0^{x^*} e^{-nf(x)}g(x) \sim \frac{e^{-nf(x_0)}}{f'(x_0)}g(x_0) \; .$$
The thesis follows from the latter result and (\ref{Jnalpha}) of Theorem \ref{thm:laplacealpha}.
\end{proof}

\end{cor}

\subsection*{A Numerical Example}

We consider the series $I(n,\alpha)$ for functions $f(x)=\frac{1}{16} - x^2 + 2 x^3 - x^4$ if $x \in[0,1]$, $+\infty$ otherwise, and
$g(x)=1$.

Functions $f,g$ satisfy the hypotheses of Theorem \ref{thm:laplacealpha}. The global minimum point
of $f$ is $\frac{1}{2}$ and $f(\frac{1}{2})=0, f''(\frac{1}{2})=1$. 

In Theorem \ref{thm:laplacealpha}, we identified 4 different kind of asymptotic behaviors of the series $I(n,\alpha)$ according to whether
$\alpha>\frac{1}{2}, \alpha=\frac12,\frac13<\alpha<\frac12,0<\alpha\leq\frac13$.

To show the different behaviors, below we plot $I(n,1)$, $I(n,\frac{1}{2})$, $I(n,\frac25)$ and $I(n,\frac{1}{3})$ against
the dominant term in the asymptotic expansion,
respectively $\sqrt{2 \pi n}$,
$\sqrt{2 \pi n } \vartheta _3\left(-\frac{1}{2} \left(\sqrt{n} \pi \right),e^{-2 \pi ^2}\right) $ ,
$n^{\frac{3}{5}}P(n,\frac{2}{5},\frac{1}{2},\frac{1}{2}) $ , and
$n^{\frac{2}{3}}P(n,\frac{1}{3},\frac{1}{2},\frac{1}{2}) $, where the function $P$ is defined by formula (\ref{eq:triangularwave}).
\begin{figure}[htpb]
  \centering
  \subfigure[The relative error $\frac{I(n,1)}{\sqrt{2 \pi n}}-1$ magnified by a factor 10 \label{fig:I1}]{\includegraphics[width=8cm,height=5.5cm]{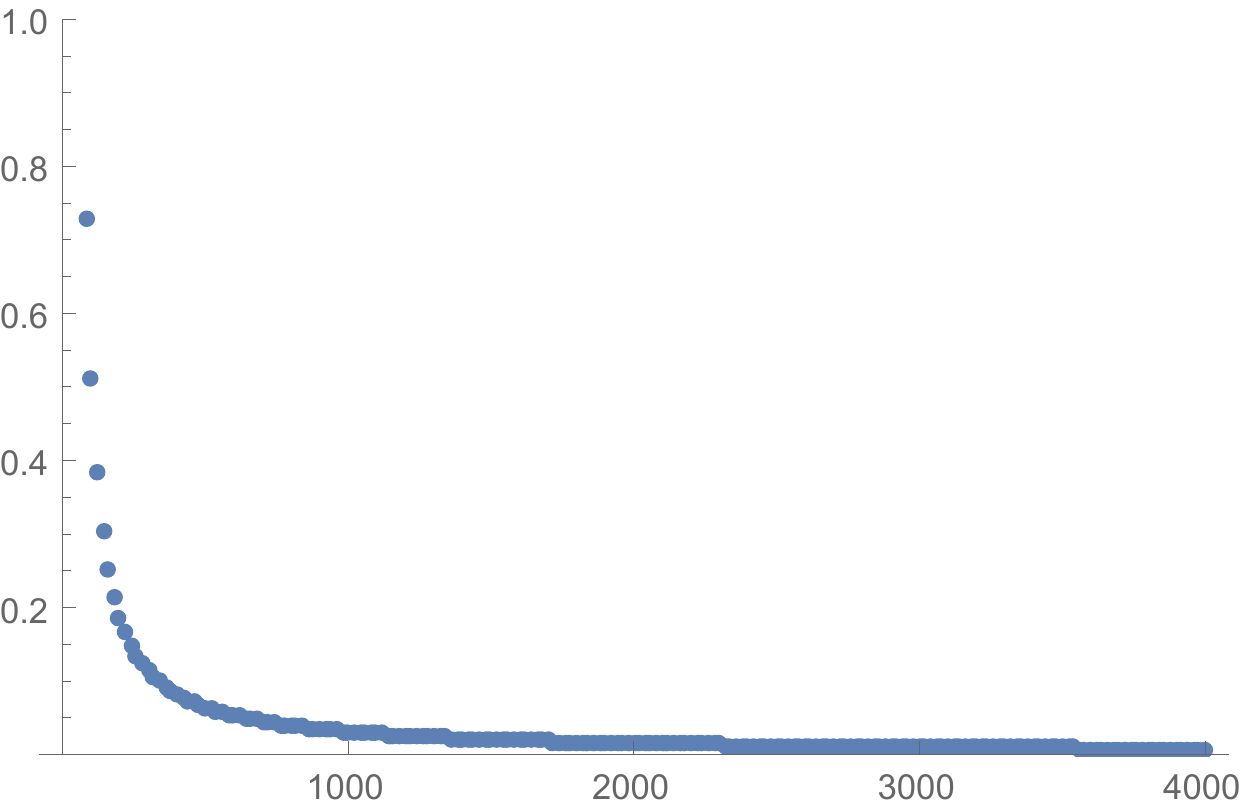}}
    \subfigure[$n^{-\frac{1}{2}}I(n,\frac{1}{2})$ in blue,
    $\sqrt{2 \pi } \vartheta _3\left(-\frac{1}{2} \left(\sqrt{n} \pi \right),e^{-2 \pi ^2}\right) $ in yellow,
    the relative error (magnified by a factor 10) in green. 
    \label{fig:I12}]{\includegraphics[width=8cm,height=5.5cm]{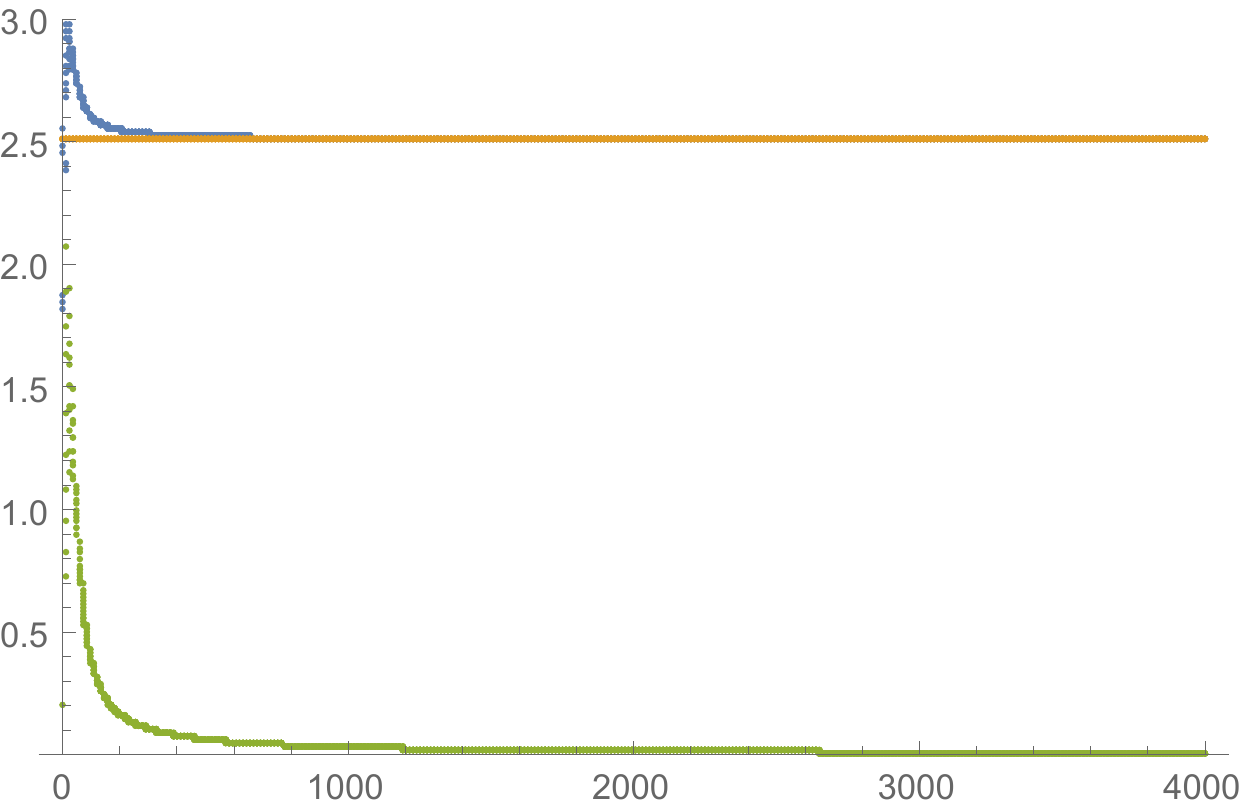}}
  \subfigure[$n^{-\frac{3}{5}}I(n,\frac{2}{5})$ in blue, $P(n,\frac{2}{5},1,\frac{1}{2})$ in yellow,
  the relative error (magnified by a factor 10) in green.
  \label{fig:I25}]{\includegraphics[width=8cm,height=5.5cm]{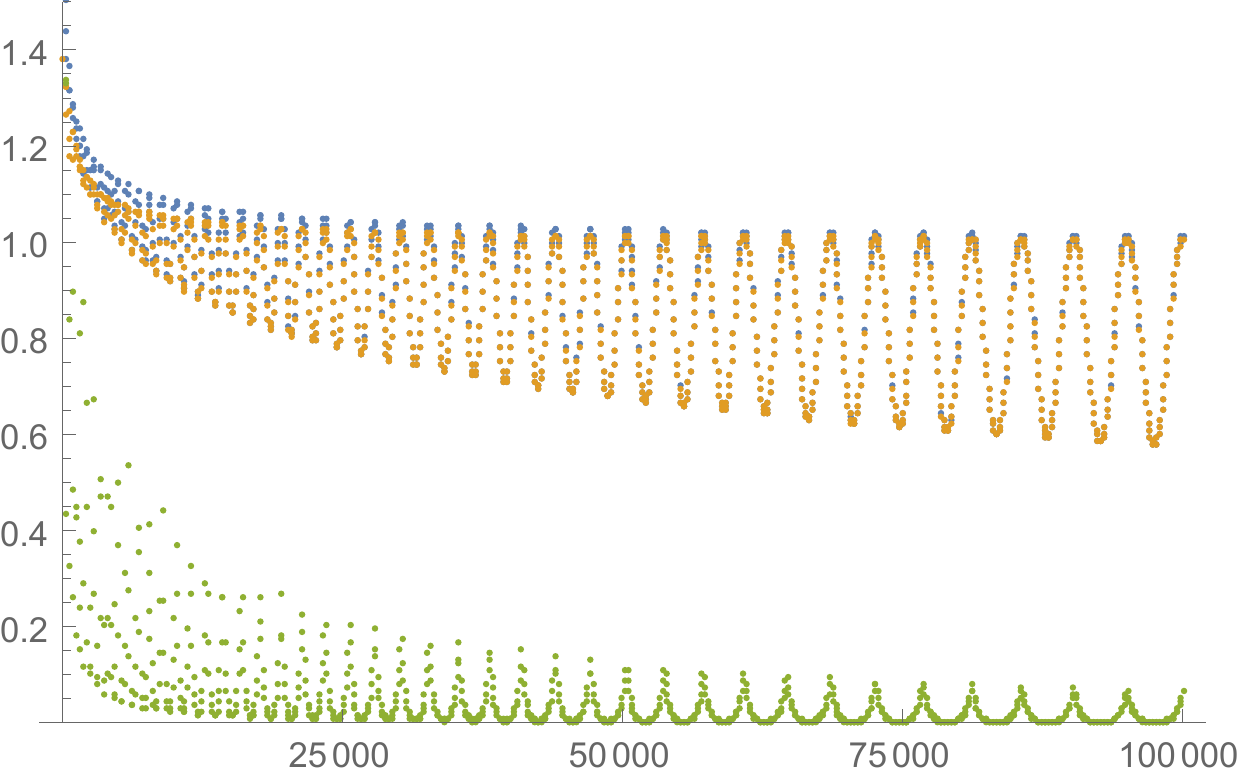}}
  \subfigure[$n^{-\frac{2}{3}}I(n,\frac{1}{3})$ in blue, $P(n,\frac{1}{3},1,\frac{1}{2})$  in yellow, relative error (magnified by a factor 100)
   in green, absolute error (magnified by a factor 100) in brown
  \label{fig:Iunterzo}]{\includegraphics[width=8cm,height=5.5cm]{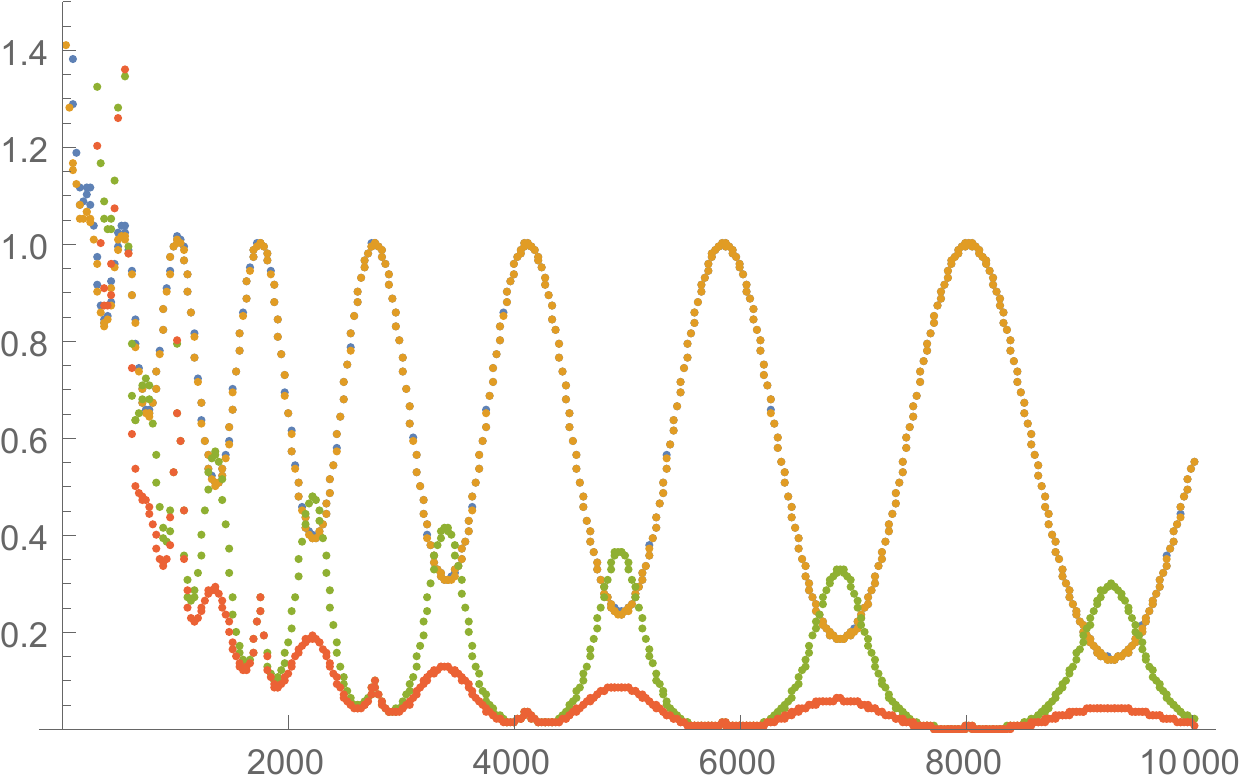}}

\end{figure}
 
From Plot \ref{fig:Iunterzo}, one can see that sequence of the relative errors  $\frac{I(n,\frac{1}{3})}{P(n,\frac{1}{3})}-1$ has a peak corresponding
to any local minimum of $P(n,\frac{1}{3})$. These peaks prevent this sequence to tend to $0$. Conversely the absolute error tends to $0$.
Both phenomena were predicted in the Corollary \ref{cor:fullasymptotic} and its proof.

\section{Semiclassical Generating Functions of SIS models}\label{section:sis}

In the present Section we apply the Laplace's method for sums to study the semiclassical limit
of models of population biology. For sake of definiteness and as a concrete example, we stick to the SIS model,
even though our considerations can be extended with few or no modifications at all to many other models of population biology admitting a
semiclassical limit, such as the ones considered in \cite{gang87} \cite{kamenev04} \cite{kamenev08} \cite{meerson10} \cite{nico13}

The SIS model is a epidemiological model of propagation of a disease.
It may be represented  by the following reaction scheme for a population of fixed size $n$
\begin{eqnarray*}
 S+I & \overset{\beta}{\to} & I+I \\
 I & \overset{\alpha}{\to} & S
\end{eqnarray*}
for the stochastic variables $I$ (infected) and $S=n-I$ (susceptible). 
Here $\beta$ is the infection rate and $\alpha$ the recovery rate. Loosely speaking, the scheme simply indicates that
a susceptible person $S$ can get infected interacting with an infected persons $I$ with rate $\beta$
while an infected person $I$ can naturally recover with rate $\alpha$.

Strictly speaking it indicates that the probability distribution $p_k$ of k individuals being infected evolves
with time according to the following linear equation (continuous time Markov chain or Master equation)
\begin{align}\label{eq:markov}
  \dot{p}_k=\frac{\beta}{N}(k-1)(N-(k-1)) p_{k-1}+\alpha(k+1)p_k-\big(\frac{\beta}{N}k(N-k)+\alpha k\big)p_{k+1} \, .
\end{align}

A simple (formal) computation \cite{nico13} shows that if we consider a semiclassical probability distribution
\begin{equation*}
 p_k \sim c_n e^{- n S(\frac{k}{n})}L(\frac{k}{n}) \, ,
\end{equation*}
then, discarding higher order contribution $O(1/n)$,
$S$ and $L$ evolves according to a pair of PDEs of classical mechanics
\begin{align} \label{eq:HJ}
& S_t+H(x,S_x)=0 \, , \; H(x,q)=  -\alpha x (e^q-1)  + \beta  x (x-1) (e^{-q}-1) \\ \label{eq:transport}
& L_t + \partial_q H(x,q)_{|q=S_x} L_x+\big( \frac{1}{2} \partial_q^2 H(x,q)_{|q=S_x} S_{xx}+\partial_q \partial_x H(x,q)_{|q=S_x}\big) L=0
\end{align}
Equation (\ref{eq:HJ}) is the Hamilton-Jacobi equation, equation (\ref{eq:transport}) is a transport equation. They both can be solved via
the method of characteristics \cite{maslov81}.

A similar (formal) computations \cite{nico13} shows that also if the generating function is semi-classical
\begin{equation*}
 \Gamma(z) = \sum_{k=0}^n p_k z^k \sim d_n e^{n \Sigma(z)}\Lambda(z)
\end{equation*}
then its evolution is described up to $O(1/n)$ correction by a (different) pair of Hamilton-Jacobi and transport equations
\begin{align} \label{eq:genHJ}
& \Sigma_t+\Theta (z,\Sigma_z)=0 \, , \; \Theta(z,q)=  -(\beta z -\alpha)(z-1) q + \beta z^2(z-1) q^2  \\ \label{eq:gentransport}
& \Lambda_t + \partial_q \Theta(z,q)_{|q=\Sigma_z} \Lambda_z+\big( \frac{1}{2} \partial_q^2 \Theta(z,q)_{|q=\Sigma_z} \Sigma_{zz}+
\beta (-z+z^2) \Sigma_z\big) \Lambda=0
\end{align}

Both approaches to the semiclassical limit of systems of population biology have been used with success within the physical literature, see e.g.
\cite{gang87} \cite{kamenev04} \cite{kamenev08} \cite{nico13};
depending on the specific problem they offer different advantages. The generating function is often useful as it seems to play the role
of the momentum-representation in quantum mechanics \cite{gang87}. Currently we are studying the long-time behavior of the semiclassical
approximation and the generating function turns out to be extremely useful to overcome the singularities that the general solution
to Hamilton-Jacobi develops \cite{nico14}.

However, there was so far no proof that the two approaches are equivalent, in the sense
that they apply to the same asymptotic regime, or in cruder words there was no proof
that a semiclassical probability distribution implies a semiclassical generating function. It was even unknown whether
the semiclassical equations, which are nonlinear, preserve the total probability.

After our Theorems on the Laplace's method for sums, we can prove that the two approaches are equivalent and that the total probability
is conserved. We can thus furnish a \textit{minimal kynematical foundation}
for the semiclassical dynamics of the SIS model, by a method that is valid for all other equation of population biology (Markov processes)
that allows a similar semiclassical limit.

Our main results are as follows
\begin{itemize}
 \item Theorem \ref{thm:generating}. First we show that if $p_k$ is semiclassical then also $\Gamma(z)$ is semiclassical and we compute explicitly $\Sigma$ and $\Lambda$.
 $\Sigma$ turns out to be the (restricted) Legendre-Fenchel transform of $S$. 
 \item Theorem \ref{cor:hamiltonjacobi}. Then we show that the $\Sigma, \Lambda$ we obtained satisfy
 the semiclassical PDEs (\ref{eq:genHJ},\ref{eq:gentransport})
 provided $S,L$ satisfy the semiclassical PDEs (\ref{eq:HJ},\ref{eq:transport}).
 \item Theorem \ref{thm:conservationofprobability}. Eventually we show, using the semiclassical dynamics of the generating function,
 that the total probability is asymptotically conserved. We prove that
 $$\sum_{k=0}^n e^{-n S(\frac{k}{n},t)}L(\frac{k}{n},t)\sim \sum_{k=0}^n e^{-n S(\frac{k}{n},0)}L(\frac{k}{n},0) $$
 if $S,L$ satisfy (\ref{eq:HJ},\ref{eq:transport}).
\end{itemize}
We warn the reader that the proofs of Theorem \ref{cor:hamiltonjacobi} and Theorem \ref{thm:conservationofprobability} require
some knowledge of the method of characteristics
for the solution of Hamilton-Jacobi equations.

\begin{thm}\label{thm:generating}
Consider the generating function of a semiclassical probability distribution
\begin{equation*}
 \Gamma(z,n)=\sum_{k=0}^n e^{-S(\frac{k}{n})}L(\frac{k}{n}) \; ,
\end{equation*}
 where 
\begin{enumerate}
 \item $S:[0,1] \to \bb{R}\cup \lbrace +\infty \rbrace$ is a continuous function, three times differentiable on $]0,1[$, and
 with a single minimum at $x_0 \in]0,1[$ and such that $S''(x) \neq 0, \forall x \in ]0,1[$. 
 \item  $L:[0,1] \to \bb{R}$ is a never vanishing differentiable function.
\end{enumerate}
 Then for any $z >0$
\begin{equation}\label{eq:S}
 \Sigma(z) \equiv \lim_{n \to \infty}\frac{\ln \Gamma(z,n)}{n}=\sup_{x \in [0,1]}\lbrace -S(x)+x \ln z \rbrace \; .
\end{equation}

Moreover, for any $z , \, e^{ S'(0)}< z <e^{S'(1)}$  let $x(z)$ be the unique solution of $f'(x)=ln z $. Then
\begin{eqnarray}\label{eq:fullasymptoticP}
\Gamma(z) \sim  e^{n \Sigma(z)}
\left\lbrace 
\begin{matrix}
 \frac{1}{1-\exp\big(-S'(1)+\log{z}\big)}  L(1) \, , \; \mbox{ if } z >e^{ S'(1)} \\
\sqrt{\frac{ \pi n}{2 S''(1)}} L(1) \, , \; \mbox{ if } z =e^{ S'(1)}\\
 \sqrt{\frac{2 \pi n}{S''(x(z))}} L(x(z)) \, , \; e^{ S'(0)}< z <e^{S'(1)} \\
\sqrt{\frac{ \pi n}{2 S''(0)}}  L(0) \, , \; \mbox{ if } z =e^{ S'(0)}\\
 \frac{1}{1-\exp\big(-S'(0)+\log{z}\big)}  L(0) \, , \; \mbox{ if } z <e^{ S'(0)}
\end{matrix} \right.
\end{eqnarray}

In particular for $z, \, e^{ S'(0)}< z <e^{S'(1)}$ the following formula holds
\begin{equation}\label{eq:compLambda}
 \Gamma(z) \sim \sqrt{2 \pi n} e^{n\Sigma(z)}\Lambda(z) \mbox{ with }\Lambda(z)=\frac{L(x(z))}{\sqrt{S''(x(z))}}
\end{equation}

\begin{proof}
For $z>0$ we can write $\Gamma(z)=\sum_k^{\infty} e^{-n\tilde{f}(\frac{k}{n})}g(\frac{k}{n})$ where
$f(x)=S(x)-x \ln z$ and $g(x)=\Lambda(x) \chi_{[0,1]}$. Here $\chi_[0,1]$ is the characteristic function of the unit interval,
namely $\chi_[0,1](x)=1$ if $x\in[0,1]$, $0$ otherwise.
The thesis follows directly from Theorems \ref{thm:laplace1} and \ref{thm:Watson1}. 

The only two cases not treated explicitly in the above mentioned Theorems are when $z =e^{S'(0)}$ or $z=e^{S'(1)}$. Here
the maximum of the summand is achieved at the boundary, where
the first derivative (but not the second) vanishes.
It is easily seen that this case is analogous to the one treated in Theorem \ref{thm:laplace1},
the only difference being a $\frac{1}{2}$ factor.
\end{proof}

\end{thm}

\begin{cor}
 With the same hypotheses of Theorem \ref{thm:generating}.
 \begin{equation}\label{eq:normalization}
\Gamma(1,n)= \sum_{k=0}^n p_k \sim \sqrt{ \frac{2\pi n}{S''(x_0)} } e^{-n S(x_0)}L(x_0)
 \end{equation}
where $x_0$ is the unique global minimum point of $S$.
\end{cor}

\begin{rem*}
Formula \ref{eq:S} shows that $\Sigma$ is the (restricted) Legendre-Fenchel transform of $S$ evaluated at $\ln z$ \cite{ilyaev03}. 
\end{rem*}

We can now prove now that $\Sigma, \Lambda$ defined as in (\ref{eq:S}, \ref{eq:compLambda}) satisfy
the semiclassical PDEs (\ref{eq:genHJ},\ref{eq:gentransport}) provided $S,L$ satisfy the semiclassical PDEs (\ref{eq:HJ},\ref{eq:transport}).

\begin{thm}\label{cor:hamiltonjacobi}
 Let $S(x,t),L(x,t) $, $t \in [0,T[ $ be solutions of equations (\ref{eq:HJ},\ref{eq:transport}) satisfying
 the hypotheses of Theorem \ref{thm:generating} for all $t$'s , and
  let $\Gamma$ be the generating functions, $\Gamma(z,t) \equiv \sum_k e^{-nS(\frac{k}{n},t)}L(\frac{k}{n},t)$.
 
 Then
 \begin{enumerate}
  \item $\Sigma(z,t)=\lim_{n \to \infty}\frac{\log \Gamma(z,t)}{n} $ exists and satisfies equations (\ref{eq:genHJ})
 for any $z>0$.
 \item For any $z$, $e^{S_x(0,t)}<z<e^{S_x(1,t)}$,
 $$\Gamma(z,t)  \sim \sqrt{2 \pi n} e^{n\Sigma(z,t)}\Lambda(z,t)  $$ where
 $\Lambda(z,t)$  satisfies (\ref{eq:gentransport}).
 \end{enumerate}

 \begin{proof}
 By Theorem \ref{thm:generating}, $\Sigma$ is the (restricted) Legendre-Fenchel transform of
 $S$ evaluated at $\ln z$. Therefore \cite{ilyaev03} letting $\tilde{\Sigma}(y,t)=\Sigma(e^y,t) $
 and $y(x,t)=S_x(x,t)$, the following identity holds
 $$
 S(x,t)+\tilde{\Sigma}(y(x,t),t)=xy(x,t) \; .
 $$
Differentiating by $t$, we get
$$
S_t+ \Sigma_t = y_t(x-\tilde{\Sigma}_y(y(x,t),t) )\; .
$$
Since $x=\tilde{\Sigma}_y(y(x,t),t)$ ($x,y$ are the conjugate variables of Legendre transform) then
$$
\tilde{\Sigma}_t(y,t)-H(\tilde{\Sigma}_y,y)=0 \; \mbox{ or equivalently } \; \Sigma_t(z,t)-H(z\Sigma_z,\ln z)=0 \; .
$$
A simple computation shows that $-H(zq,\ln z)=\Theta(z,q)$, for any $z,q$.
 
A similar, but more involved, computation shows that  $\Lambda$ satisfies the transport equation.
Details can be found, for example, in the proof of Theorem 5.3 in \cite{maslov81}.
We omit to reproduce here those computations since they require much more machinery from
Hamilton-Jacobi theory than the one appropriate for the present paper. 
 \end{proof}

\end{thm}

Using the semiclassical equations for $\Gamma(z)$, we can show that the semiclassical dynamics conserves the total probability.

\begin{thm}\label{thm:conservationofprobability}
 Let $S(x,t),L(x,t) $, $t \in [0,T[ $ be solutions of equations (\ref{eq:HJ},\ref{eq:transport}) satisfying
 the hypotheses of Theorem \ref{thm:generating} for all $t$'s.
 Then
 \begin{equation}
 \sum_{k=0}^n e^{-n S(\frac{k}{n},t)}L(\frac{k}{n},t)\sim \sum_{k=0}^n e^{-n S(\frac{k}{n},0)}L(\frac{k}{n},0) \, , \; \forall t \in [0,T[  \; .
 \end{equation}
\begin{proof}
 After Theorem \ref{thm:generating}, we have
 $\sum_{k=0}^n e^{-n S(\frac{k}{n},t)}L(\frac{k}{n},t) \sim e^{n \Sigma(1,t)}\Lambda(1,t)$.
 To prove the thesis
 it is sufficient to show that
 $\Sigma(1,t)$ and $\Lambda(1,t)$ are constant in time. We do that by means of the method of characteristics:

 On the solution of the system of ODEs 
 $$\dot{z}=\partial_q \Theta(z,q), \dot{q}=-\partial_z \Theta(z,q) , \, q(0)=\Sigma_z(z,0)$$ then
\begin{align*}
 & \Sigma(z(t))=\Sigma(z(0))+ \int_0^t\partial_q \Theta(z(s),q(s)) q(s) ds \\
 & \Lambda(z(s))=\Lambda(z(0))-\int_0^t\big( \frac{1}{2} \partial_q^2 \Theta(z(s),q(s)) \Sigma_{zz}(z(s))+
\beta (-z(s)+z^2(s)) \Sigma_z(z(s)) \big) ds \; .
\end{align*}
Since $\dot{z}=0 $ if $z=0$ then the characteristic starting at $z=0$ will stay in $z=0$ for all $t$.
Moreover, the integrals defining $S,L$ vanish for all $t$ as the integrands vanish identically at $z=0$. Therefore
 $\Sigma(1,t)=\Sigma(1,0), \Lambda(1,t)=\Lambda(1,0)$.
\end{proof}

\end{thm}

\begin{rem*}
We can actually relax the hypotheses of Theorem \ref{thm:conservationofprobability}. Indeed it holds, with a slightly more complex proof,
if we just requires that $S''(x,0)$ and $L(x,0)$ do not vanish in a neighborhood of the global minimum of $S(x,0)$. 
\end{rem*}

\subsection*{An apparent paradox concerning the meta-stable state of SIS model}
It has been noticed in many models of population biology, see e.g.  \cite{meerson10,kessler07,nico13},
that the in the semiclassical regime the meta-stable states of the model correspond to non-trivial
stationary solutions of the Hamilton-Jacobi equation.

Assuming $\frac{\beta}{\alpha}>0$, the SIS model admits a quasi-stationary or meta-stable state $p^*$, corresponding
to an exponentially small eigenvalue of the transition matrix of the Markov chain  (\ref{eq:markov}),
see \cite{nico13}.

\begin{figure}[htpb]
  \centering
  \subfigure[Level curves of $H(x,q)$. The thick line is the derivative $q^*(x)$ of the non-trivial stationary solution
   (\ref{eq:sisstationary}).
   \label{fig:stationaryprob}]
  {\includegraphics[width=5cm]{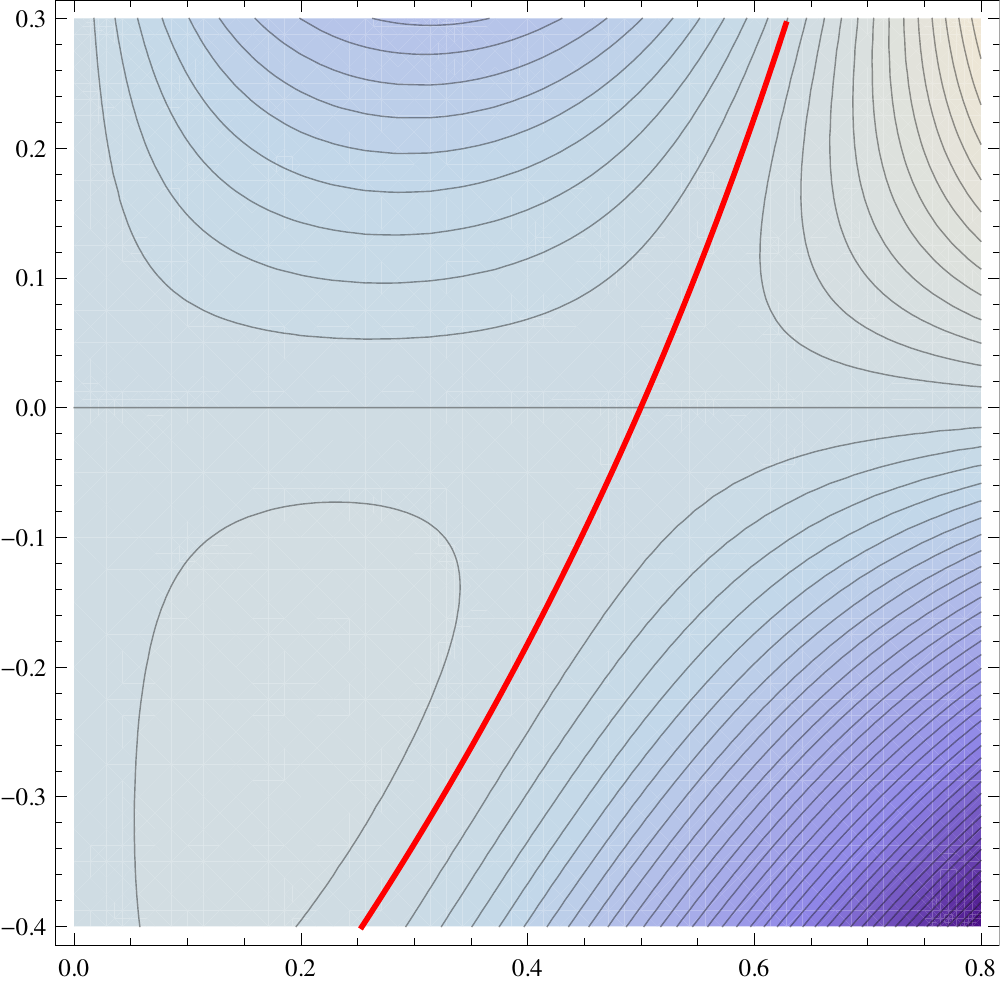}}
  \subfigure[Level curves of $\Theta(z,q)$. The thick line is the derivative of the unphysical stationary solution.
  \label{fig:stationarygenwrong}]{\includegraphics[width=5cm]{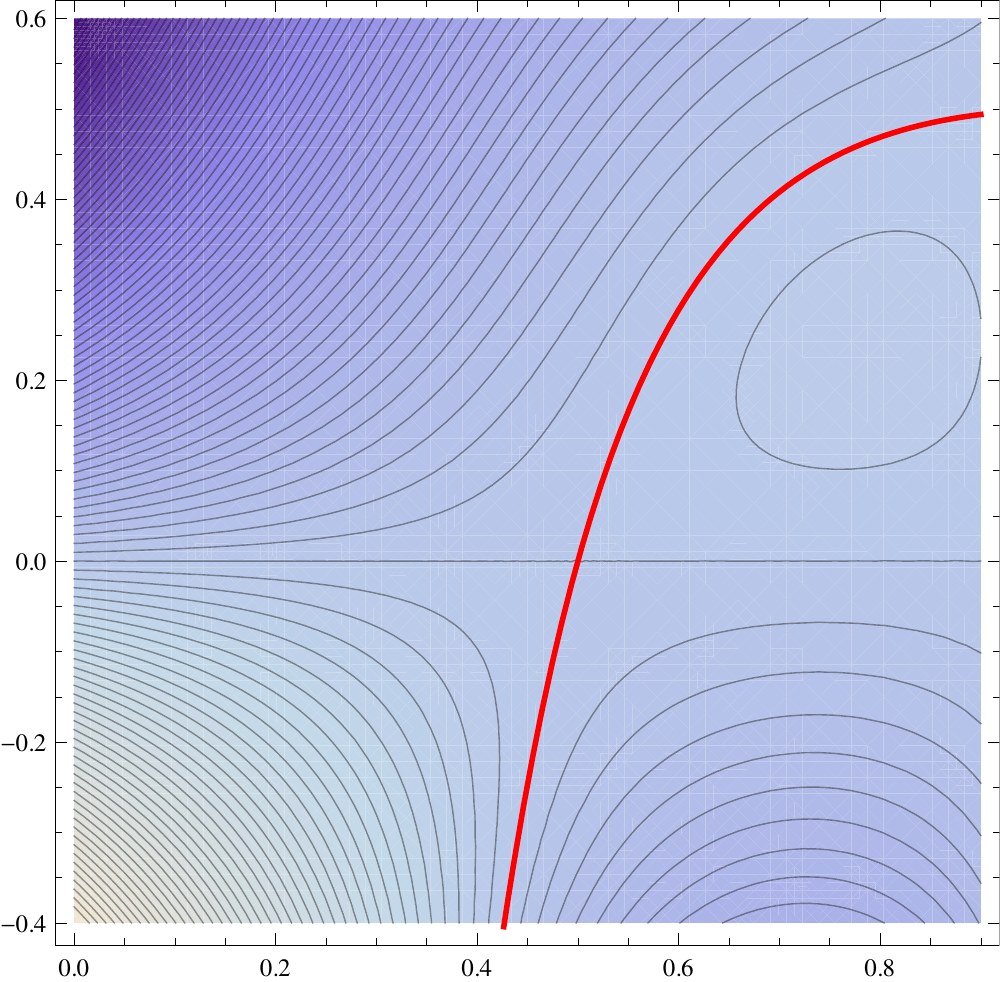}}
  \subfigure[Level curves of $\Theta(z,q)$. The thick line is the derivative of the correct (non-smooth) stationary
  solution $Q^*(z)$ (\ref{eq:sisstationarygen}). \label{fig:stationarygen}]
  {\includegraphics[width=5cm]{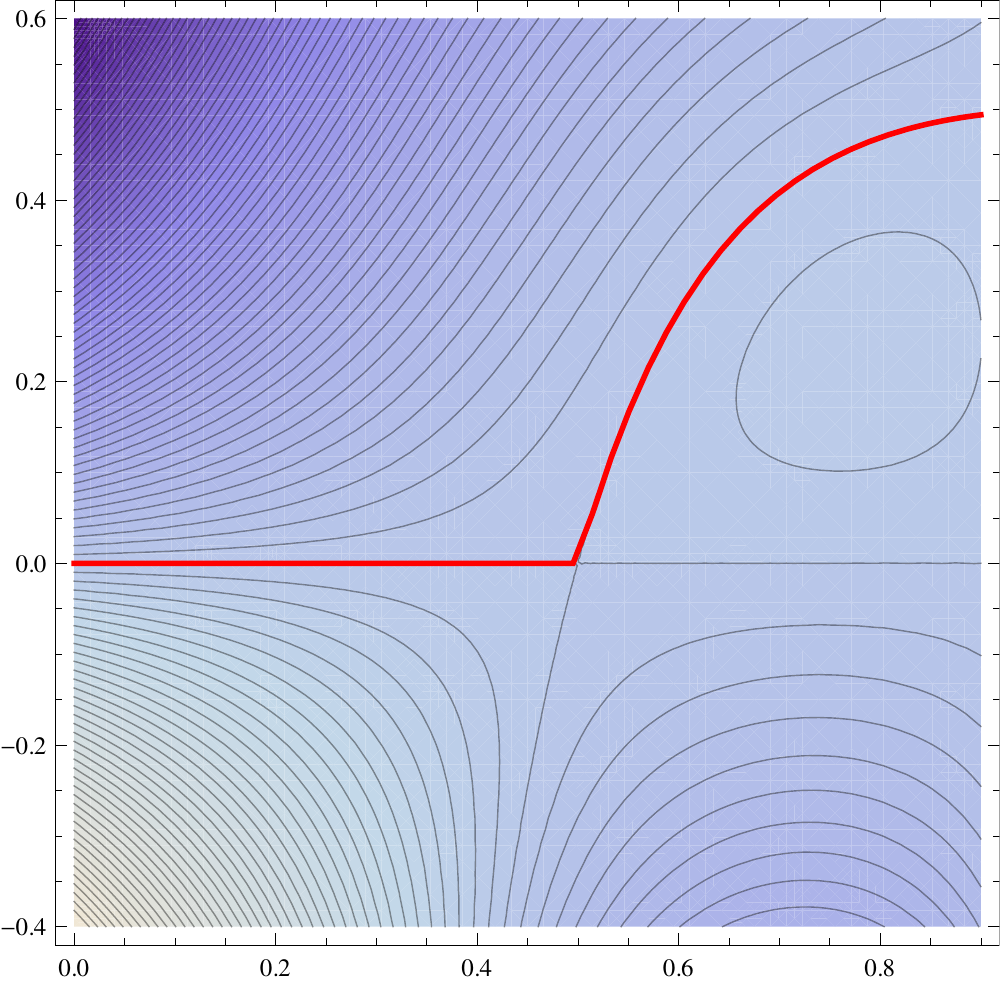}}
\end{figure}

In the semiclassical regime and considering the probability distribution
picture, the meta-stable state corresponds to a non-trivial stationary solution $S^*$ of the Hamilton-Jacobi equation (\ref{eq:HJ}).
This in turn corresponds to a zero-energy level-curve of the Hamiltonian as
clearly $S^*_t=0$ if and only if $H(x,S_x^*(x))=0$.
Explicitly \cite{nico13} (see Figure \ref{fig:stationaryprob})
\begin{equation}\label{eq:sisstationary}
 S^*(x)=q_x^*(x) \, , \;    q^*(x) =\log \frac{\beta(1-x) }{\alpha }
\end{equation}

Surprisingly, in the generating function picture such meta-stable state seems to be missing.
In fact, since for any probability distribution $p_k \geq 0 , \forall k$ then meta-stable state should correspond to
zero-energy level-curves of
the Hamiltonian $\Theta(z,Q^*(z))$ such that $Q^*(z) \geq 0$. However,
the only non-trivial smooth energy-level curve of the Hamiltonian is $ \frac{-\alpha+\beta z}{\beta z^2} $ that
tends to $-\infty$ as $z \to 0^+$, see Figure \ref{fig:stationarygenwrong}.
This seems to contradict Theorem \ref{cor:hamiltonjacobi}, after which any stationary solution to
(\ref{eq:HJ}) corresponds to a stationary solution
of (\ref{eq:genHJ}).

The solution of this apparent paradox comes from Theorem \ref{thm:generating}. After this Theorem,
we know that $\Sigma^*$ is not smooth even if $S^* $ is smooth. Indeed the second derivative of
$\Sigma^*$ is discontinuous at $z=e^{q^*(0)}, z=e^{-q^*(1)}$ (in this case $e^{-q^*(1)} =+\infty $).
This is because $\Sigma^*$ is the restricted Legendre-Fenchel transform of $S^*$,
formula (\ref{eq:S}). In fact, given $S^*$ as above and using formula (\ref{eq:S}), we obtain
\begin{equation}\label{eq:sisstationarygen}
 \Sigma^*(z)=Q_z^{*}(z) \, , \;  Q^*(z) = 0 \mbox{ if } z <\frac{\alpha}{\beta} , \frac{-\alpha+\beta z}{\beta z^2} \mbox{ otherwise}.
\end{equation}
$Q^*(z)$ is thus the union of two branches of two different smooth zero-energy level-curves of $\Theta(z,q)$, the
trivial curve for $z < \frac{\beta}{\alpha}$ and the non-trivial one for $ z\geq  \frac{\beta}{\alpha} $, 
see Figure \ref{fig:stationarygen}.

\section{Concluding Remarks}

We have analyzed the series $I(n,\alpha)$ and shown that its asymptotic behavior can be effectively computed for all values of $\alpha$.
Most of the proofs we have given are based on the comparison of the given series with a standard exponential or Gaussian series.
On the other hand, we showed in the proof of Theorem \ref{thm:laplace1} that it is possible to compute the asymptotic behavior of
the series by comparing it with the integral ${\cal{I}}(n)$. We did that by means of the Euler-McLaurin summation formula. As a consequence,
we obtained two different ways of dealing with the series $I(n,\alpha)$.

The same methods we use allow to deal with a number of variations of the original problem. For example,
in this work we have chosen to consider only functions $f$ such that the global minimum is not-degenerate,
that is either $f'(0)>0$ or $f''(x_0)>0$. However, degenerate cases can be easily dealt along the lines of this work.

Also multidimensional series like
\begin{equation*}
 \sum_{k_1,\dots, k_m}e^{- n f(\frac{k_1}{n},\dots,\frac{k_m}{n})}g(\frac{k_1}{n},\dots,\frac{k_m}{n})
\end{equation*}
are amenable to the same analysis.

For what concerns the application to the semiclassical limit of population biology, we stick to the SIS model. As it was already noticed,
our results in this respect can be extended with few or no modifications at all to many other models of population biology admitting
a semiclassical limit.

In the impossibility of tackling the seemingly endless possible variations of the Laplace's method and its applications,
we believe we have furnished the interested reader enough instruments
to tackle some of the possible generalizations

We conclude the paper with few interesting open problems.

The first problem is the asymptotic behaviour of the generating function for $z$ not real and positive.
For any $f,g$, the generating function  can be cast into a series of the kind $I(n,\alpha=1)$
\begin{equation*}
 \Gamma(z)=\sum_{k=0}^n e^{-n \tilde{f}(\frac{k}{n})}g(\frac{k}{n}) \, , \;  \tilde{f}(x)=f(x)- x (\ln |z| +i \arg z) 
\end{equation*}
but if $z$ is not positive, $\tilde{f}$ is not real.
Since the leading contributions are the ones
close to the global maximum of $|e^{-nf(\frac{k}{n})}|$, that is close to the minimum point $x(z)$ of $Re\tilde{f}$, we are led to
approximate $\Gamma(z)$ by the following Gaussian series with an imaginary linear term
$$
e^{-n\tilde{f}(x(z))}\sum_k e^{-\frac{f''(x(z))}{2}\frac{ (k- nx(z))^2 +  i n \arg{z} k}{n}} \sim e^{-nf(x(z))}
e^{- \frac{n}{2 f''(x(z))} \arg^2 z+ i n \arg z x(z)} \; .
$$
The latter result stems from the Poisson summation formula applied to the series.

If $\arg z$ is not real and positive, the latter series is thus exponentially small
and even smaller than (the estimate of) 
the error term generated by neglecting contribution outside the maximum of $|e^{-n\tilde{f}(\frac{k}{n})}|$,
which is - after the proof of Theorem \ref{thm:laplacealpha}- 
$e^{-n\tilde{f}(x(z))} O(e^{-\frac{f''(x_0)}{2}} n^{1-\beta})$, for any $0<\beta <\frac{1}{2}$.
Therefore we can conclude that $\Gamma(z)$ is exponentially small relative to $\Gamma(|z|)$ if $z \neq |z|$,
but we do not know its precise asymptotic behavior.

The second problem, which is related to the first one in case $f,g$ are analytic, is the steepest-descent method for the series $I(n,1)$.
Let in fact $f,g$ be analytic (not real) functions and suppose $[0,\infty[$ can be deformed into a path of steepest descent for $f$. We
can then compute
${\cal{I}}(n,1)$ by the method of steepest descent.
However, does $I(n) \sim {\cal{I}}(n)$ if ${\cal{I}}(n) $ is computed by deforming the integration path?
Unfortunately, the Euler-McLaurin formula (\ref{eq:eulermclaurin}) we use to estimate the difference between the series and the integral,
does not allow to consider deformations of the integration contour
because the Bernoulli functions are \textbf{not} analytic.

Finally the most relevant open problem concerns the global-in-time semiclassical analysis of epidemiological models.
This issue has been recently addressed in \cite{nico14} for some particular cases, but it still lacks a general solution.
The obstacle to a global-in-time semiclassical analysis arises from the solution of the Hamilton-Jacobi equation (\ref{eq:HJ}). In fact,
for quite general initial data the Hamilton-Jacobi equation develops a singularity in a finite time after which classical
solutions become multi-valued. Therefore after this time the probability distribution cannot be described in the
form $p_k \sim e^{-nS(\frac{k}{n},t)}L(\frac{k}{n},t)$ for some smooth solutions $S,L$ of the Hamilton-Jacobi and transport equation
(\ref{eq:HJ},\ref{eq:transport}).
In Quantum Mechanics this problem can be overcome \cite{maslov81} by considering the superposition of the different
classical solutions, namely a wave function of the form
$$
\Psi(x,\hbar) \sim \sum_k e^{i \frac{\varphi_k + S_k(x)}{\hbar}}L_k(x) \, , \; \hbar \to 0
$$
where $S_k$ are the different values of $S$ at $x$ and $\varphi_k$ some locally constant phases.

However, the superposition principle does not apply in the case of epidemiological models simply because
$\sum_{k}e^{-nS_k(k/n)}L(k/n) \sim  e^{-nS_j(k/n)}L_j(k/n)$ where $S_j=\inf S_k$.
This corresponds to the fact that the underlying differential equations
(the Markov chain) are essentially dissipative. We therefore expect that the global-in-time asymptotics of the probability
distribution can be given in terms of a dissipative regularization of the Hamilton-Jacobi equation (\ref{eq:HJ}).

\end{document}